\documentclass[12pt]{article}
\usepackage{bbm}
\usepackage{mathrsfs}
\usepackage{amsfonts}
\usepackage{float}
\usepackage{amsthm}
\usepackage{epsfig,rotating}
\usepackage{amssymb,amsmath}
\usepackage{latexsym}
\usepackage{graphicx,subfigure}
\usepackage{multirow}
\usepackage{booktabs,natbib}
\usepackage[]{algorithm2e}
\usepackage{subfigure}
\usepackage{comment}
\usepackage{natbib}
\usepackage{xcolor}

\newcommand{\be}{\begin{equation}}
\newcommand{\ee}{\end{equation}}
\newcommand{\beaa}{\begin{eqnarray*}}
\newcommand{\eeaa}{\end{eqnarray*}}
\newcommand{\bea}{\begin{eqnarray}}
\newcommand{\eea}{\end{eqnarray}}
\newcommand{\bal}{\begin{align}}
\newcommand{\eal}{\end{align}}
\newcommand{\bali}{\begin{align*}}
\newcommand{\eali}{\end{align*}}

\newcommand{\la}{\langle}
\newcommand{\ra}{\rangle}

\newcommand{\bbx}{{\bf {x}}}
\newcommand{\bby}{{\bf {y}}}
\newcommand{\bbz}{{\bf {z}}}
\newcommand{\bbv}{{\bf {v}}}
\newcommand{\bbu}{{\bf {u}}}

\newcommand{\bbX}{{\bf {X}}}

\newcommand{\bbZ}{{\bf Z}}

\newcommand{\bbA}{{\bf A}}
\newcommand{\bbB}{{\bf B}}

\newcommand{\mR}{\mathbb{R}}
\newcommand{\mE}{\mathbb{E}}
\newcommand{\mP}{\mathbb{P}}

\newcommand{\prk}{{R(\boldsymbol{\beta})}}
\newcommand{\erk}{{\widehat{R}_n(\boldsymbol{\beta})}}
\newcommand{\bbeta}{\boldsymbol{\beta}}
\newcommand{\bxi}{\boldsymbol{\xi}}
\newcommand{\bmu}{\boldsymbol{\mu}}

\newcommand{\bGamma}{\boldsymbol{\Gamma}}
\newcommand{\bSigma}{\boldsymbol{\Sigma}}
\newcommand{\hbeta}{\boldsymbol{\hat\beta}}
\newcommand{\tbeta}{\boldsymbol{\tilde\beta}}

\newcommand{\gbeta}{G(\boldsymbol{\beta})}
\newcommand{\jbeta}{D_G(\boldsymbol{\beta})}
\newcommand{\Tbeta}{D^2_G(\boldsymbol{\beta})}
\newcommand{\dbeta}{D^3_G(\boldsymbol{\beta})}

\newcommand{\ball}{{\bf B}^p(r)}
\newcommand{\sball}{{\bf B}^p(\boldsymbol\tbeta^{\star},\varepsilon_0)}
\newcommand{\gtbeta}{G(\boldsymbol{\tilde\beta^{\star}})}

\newcommand{\sbeta}{\boldsymbol{\beta}^{\star}}
\newcommand{\tsbeta}{\boldsymbol{\tilde\beta}^{\star}}
\newcommand{\hobeta}{\boldsymbol{\hat\beta}^{\mathcal{O}}}
\newcommand{\thrbeta}{\boldsymbol{\hat\beta}^{\text{thr}}}

\newcommand{\oball}{{\bf B}^p_0(s_0)}

\newtheorem{assumption}{\noindent A{\footnotesize SSUMPTION}}
\newtheorem{theorem}{\noindent T{\footnotesize HEOREM}}
\newtheorem{prop}{\noindent P{\footnotesize ROPOSITION}}
\newtheorem{lemma}{\noindent L{\footnotesize EMMA}}

\newtheorem{remark}{\noindent R{\footnotesize EMARK}}

\providecommand{\keywords}[1]
{
	\small	
	\textbf{\textit{Keywords---}} #1
}

\title{Robust High-Dimensional Regression with Coefficient Thresholding and its Application to Imaging Data Analysis
}
\author{
Bingyuan Liu$^1$, Qi Zhang$^1$, Lingzhou Xue$^1$, Peter X.-K. Song$^2$, and Jian Kang$^2$\\ $^1$Pennsylvania State University and $^2$University of Michigan
}
\date{First Version: November 2019; \\ This Version: January 2021}
\begin{document}

\maketitle

\begin{abstract}
It is of importance to develop statistical techniques to analyze high-dimensional data in the presence of both complex dependence and possible outliers in real-world applications such as  imaging data analyses. We propose a new robust high-dimensional regression with coefficient thresholding, in which an efficient nonconvex estimation procedure is proposed through a thresholding function and the robust Huber loss. The proposed regularization method accounts for complex dependence structures in predictors and is robust against outliers in outcomes. Theoretically, we  analyze rigorously the landscape of the population and empirical risk functions  for the proposed method. The fine landscape enables us to establish both {statistical consistency and computational convergence} under the high-dimensional setting. Finite-sample properties of the proposed method are examined by extensive simulation studies. An illustration of real-world application concerns a scalar-on-image regression analysis for an association of psychiatric disorder measured by the general factor of psychopathology with features extracted from the task functional magnetic resonance imaging data in the Adolescent Brain Cognitive Development study.
\end{abstract}

\keywords{Landscape analysis; Nonconvex optimization; Thresholding function; Scalar-on-image regression.}

\section{Introduction}

Regression analysis of high-dimensional data has been extensively studied in a number of research fields over the last three decades or so. To overcome the high-dimensionality, statistical researchers have proposed a variety of regularization methods to perform variable selection and parameter estimation simultaneously. Among these, the $\ell_0$ regularization enjoys the oracle risk inequality \citep{barron1999risk} but it is impractical due to its NP-hard computational complexity. In contrast, the $\ell_1$ regularization \citep{tibshirani1996regression} provides an effective convex relaxation of the $\ell_0$ regularization and achieves variable selection consistency under the so-called \emph{irrepresentable condition} \citep{zhao2006model,zou2006adaptive,wainwright2009sharp}. The adaptive $\ell_1$ regularization \citep{zou2006adaptive} and the folded concave regularization \citep{fan2001variable,zhang2010nearly} relax the \emph{irrepresentable condition} and improve the estimation and variable selection performance. The folded concave penalized estimation can be implemented through a sequence of the adaptive $\ell_1$ penalized estimations and achieves the strong oracle property \citep{zou2008one, fan2014strong}. 

\begin{figure}[H]\label{illustration}
	\centering
	\subfigure[]{\includegraphics[width=2in]{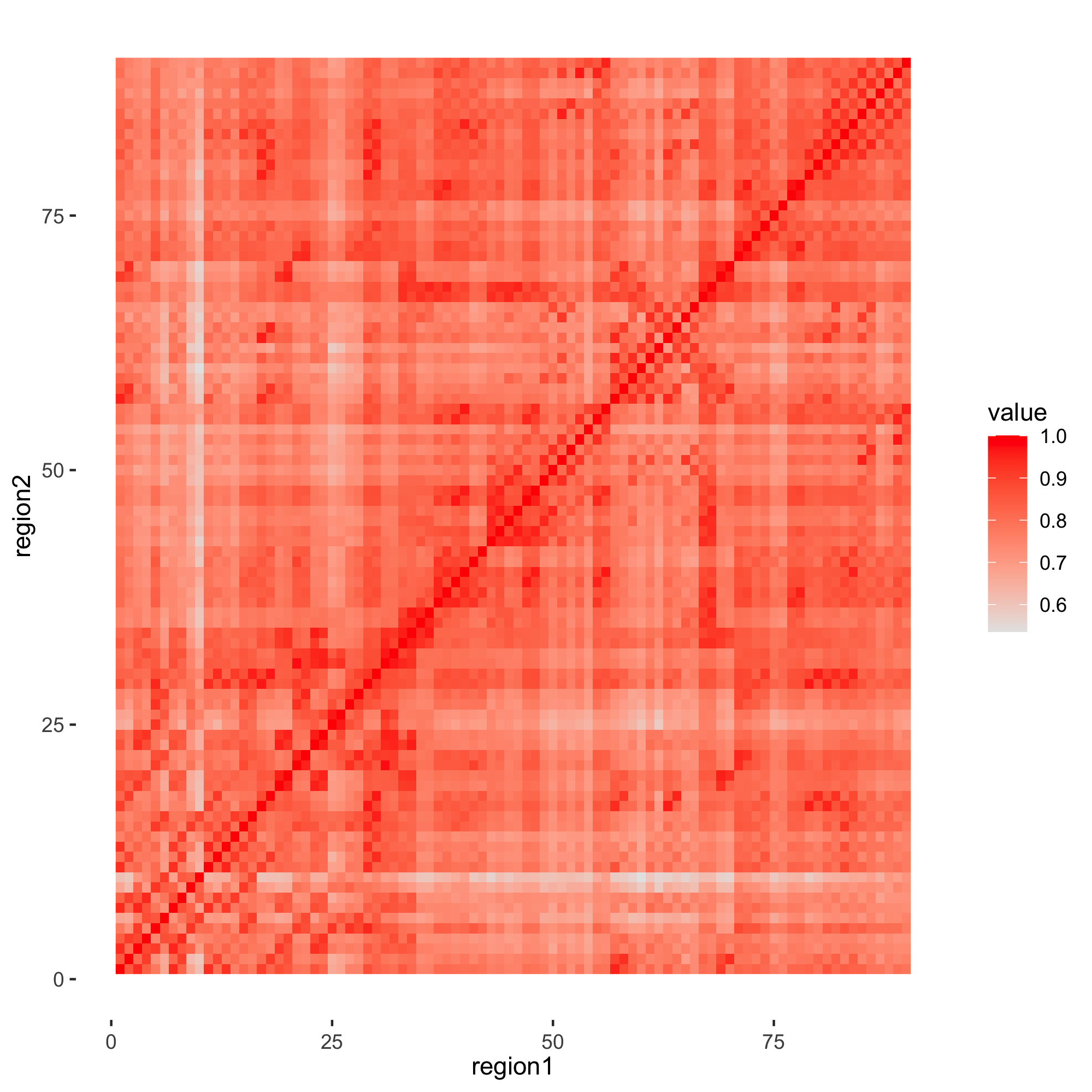} }
	\subfigure[]{\includegraphics[width=2in]{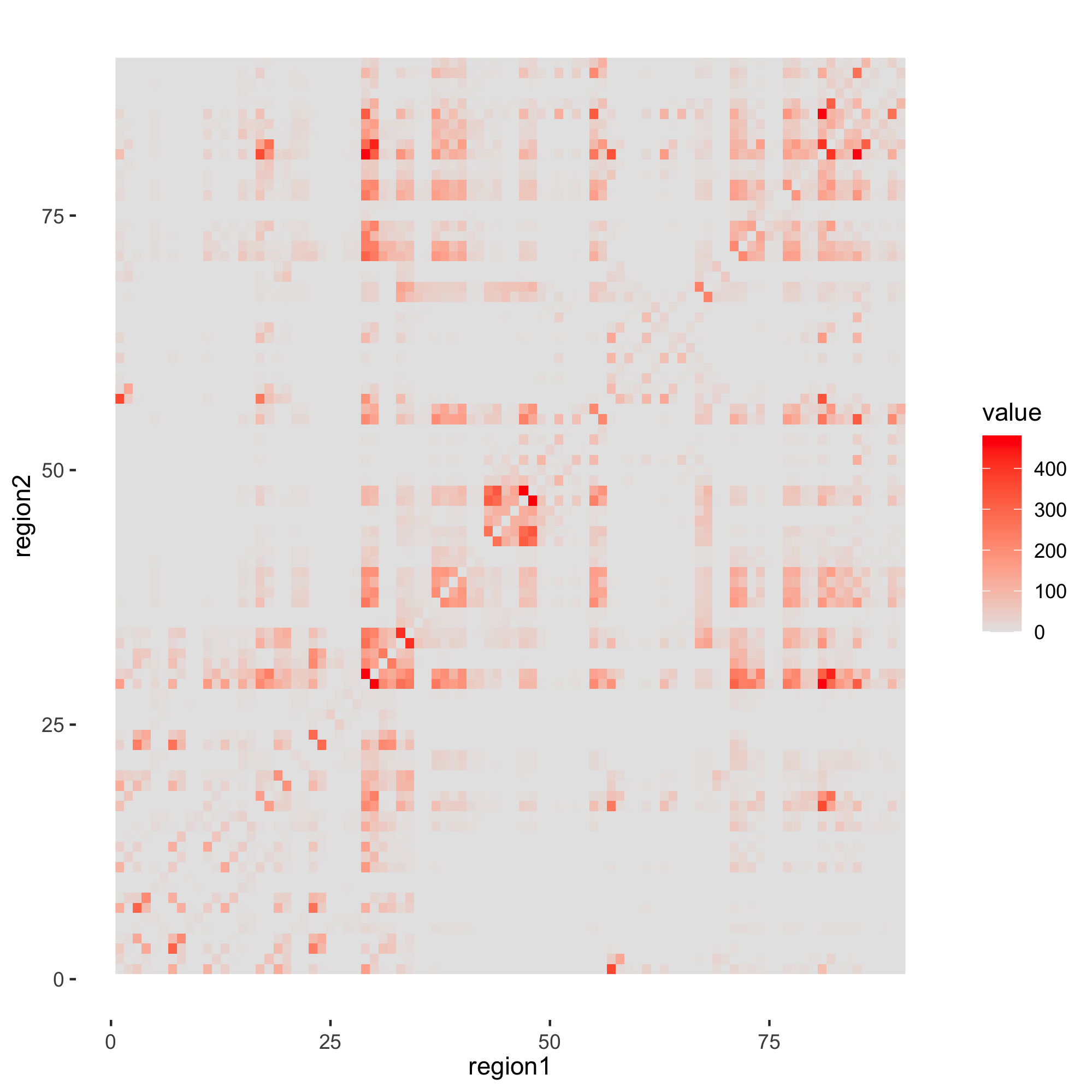} }
	\caption{Illustration of the strong dependence structure among imaging predictors. Panel (a)  shows the largest absolute value of correlation coefficients between  regions, and Panel (b) shows the number of correlation coefficients that are $\ge 0.8$ or $\le -0.8$ between  regions.}\label{cor_illustration}
\end{figure}

Despite these important advances, existing methods, including the (adaptive) $\ell_1$ regularization and folded concave regularization, do not work well when predictors are strongly correlated, which is the case especially in scalar-on-image regression analysis \citep{wang2017generalized,kang2018scalar,heselective}. This paper is motivated by the needs of analyzing the $n$-back working memory task fMRI data in the Adolescent Brain Cognitive Development~(ABCD) study~\citep{casey2018adolescent}. The task-invoked fMRI imaging measures the blood oxygen level signal that is linked to personal neural activities when performing a specific task. The $n$-back task is a commonly used approach to making assessment in psychology and cognitive neuroscience with a focus on working memory. One question of interest is to understand the association between  the risk of developing psychiatry disorder and features related to functional brain activity. We use the 2-back versus 0-back contrast map statistics derived from the $n$-back task fMRI data as imaging predictors. We aim at identifying  important  imaging biomarkers that are strongly associated with the  general factor of psychopathology~(GFP) or ``p-factor".
GFP is a psychiatric disorder outcome used to  evaluate the overall mental health of a subject. In this application, it is expected that the \emph{irrepresentable condition} of the $\ell_1$ regularization can be easily violated by strong dependence among high dimensional imaging predictors from fMRI data. To illustrate the presence of strong dependence among imaging predictors, in Figure \ref{illustration}, we plot the largest absolute value of correlation coefficients and the number of correlation coefficients that are $\ge 0.8$ or $\le -0.8$ between brain regions. Obviously, there exists strong dependence among imaging predictors, so that existing methods may not have a satisfactory performance in the scalar-on-image analysis. {See the simulation study in Section 4 for more details.}

To effectively address potential technical challenges in the presence of such strongly correlated predictors, we consider a new approach based on the thresholding function technique. The rationale behind our idea is rooted in attractive performances given by various recently developed thresholding methods, including the \emph{hard-thresholding property} of the $\ell_0$ regularization shown in \cite{fan2013asymptotic2}. They showed that the global minimizer in the thresholded parameter space enjoys the variable selection consistency. Thus, with proper thresholding of coefficients, it is possible to significantly relax the irrepresentable condition while to address the strong dependence in the scalar-on-image regression analysis. Recently, manifested by the potential power of the thresholding strategy, \cite{kang2018scalar} studied a new class of Bayesian nonparametric models based on the soft-thresholded Gaussian prior, and \cite{sun2019hard} proposed a two-stage hard thresholding regression analysis that applies a hard thresholding function on the initial $\ell_1$-penalized estimator. To achieve the strong oracle property \citep{fan2014strong}, \cite{sun2019hard} required the initial solution is reasonably close to the true solution in aspect of $\ell_2$ norm with high probability.

Robustness against outliers occurring from heavy-tailed errors is of great importance in the scalar-on-image analysis. Due to various limitations of used instruments and quality control in data preprocessing, fMRI data often involves many potential outliers\citep{poldrack2012future}, compromising the stability and reliability of standard regression analyses. 
fMRI indirectly measures neural activity by assessing blood-oxgen-level-dependent signals and its signal-to-noise ratio is often low\citep{lindquist2008statistical}. Moreover, the complexity of fMRI techniques limits the capacity of  unifying fMRI data preprocessing procedures~\citep{bennett2010reliable,brown2017controversy} to identify and remove outliers effectively. Standard regression analysis with contaminated data 
may lead to a high rate of false positives in inference, as shown in many empirical studies~\citep{eklund2012does,eklund2016cluster}. It is loudly advocated that potential outliers should be taken into account in the study of brain functional connectivity using fMRI data \citep{rosenberg2016neuromarker}. In the lens of robustness, many works have been proposed to study the high dimensional robust regression problem. \cite{el2013robust} studied the consistency of regression with a robust loss function such as the least absolute deviation (LAD).  In a high-dimensional robust regression, \cite{loh2017statistical} showed that 
the use of a robust loss can help achieve the optimal rate of regression coefficient estimation with independent zero-mean error terms. In addition, \cite{loh2018scale} showed that by calibrating with a scale estimator in the Huber loss, the regularized robust regression estimator can be further improved. 

In the current literature of the high dimensional scalar-on-image regression, \cite{goldsmith2014smooth} introduced a single-site Gibbs sampler that incorporates spatial information in a Bayesian regression framework to perform the scalar-on-image regression. \cite{li2015spatial} introduced a joint Ising and Dirichlet process prior to develop a Bayesian stochastic search variable selection.  \cite{wang2017generalized} proposed a generalized regression model in which the image is assumed to belong to the space of bounded total variation incorporating the piece-wise smooth nature of fMRI data. 
Motivated by these works, in this paper we first introduce a new integrated robust regression model with coefficient thresholding and then propose a penalized estimation procedure with provable theoretical guarantees, where the noise distribution is not restricted to be sub-Gaussian.  Specifically, we propose to use a smooth thresholding function to approximate the discrete hard thresholding function and tackle the strong dependence of predictors together with the use of the smoothed Huber loss \citep{charbonnier1997deterministic} to achieve desirable robust estimation. We design a customized composite gradient descent algorithm to efficiently solve the nonconvex and nonsmooth optimization problem. The proposed  coefficient thresholding method is capable of incorporating intrinsic group structures of high-dimensional imaging predictors and dealing with their strong spatial and functional dependencies. Moreover, the proposed method effectively improves robustness and reliability.  

The proposed regression with the coefficient thresholding method results in a nonconvex objective function in optimization. In the current literature, it becomes an increasingly important research topic to obtain the statistical and computational guarantees for nonconvex optimization methods. 
The local linear approximation (LLA) approach \citep{zou2008one,fan2014strong,fan2018lamm} and the Wirtinger flow method \citep{candes2015phase,cai2016optimal} directly have enabled to analyze the computed local solution. The restricted strong convexity (RSC) condition \citep{negahban2009unified,negahban2012restricted,loh2013regularized,loh2017support} and the null consistency condition  \citep{zhang2012general} were used to prove the uniqueness of the sparse local solution. However, it still remains non-trivial to study theoretical properties of the proposed robust regression with coefficient thresholding. The nonconvex optimization cannot be directly solved by the LLA approach, and the RSC condition does not hold. Alternatively, following the seminal paper by \cite{mei2016landscape}, we carefully study the landscape of the proposed method. We prove that the proposed nonconvex loss function has a fine landscape with high probability and also establish the uniform convergence of the directional gradient and restricted Hessian of the empirical risk function to their population counterparts. Thus, under some mild conditions, we can establish key statistical and computational guarantees. Let $n$ be the sample size, $p$ be the dimension of predictors, and $s$ the size of the support set of true parameters. Specifically, we prove that, with high probability, (i) any stationary solution achieves the oracle  inequality under the $\ell_2$ norm when $n \ge C s\log p$; (ii) the proposed nonconvex estimation procedure has a unique stationary solution that is also the global solution when $n \ge C s\log^2 p$; and (iii) the proposed composite gradient descent algorithm attains the desired stationary solution. {We shall point out that both statistical and computational guarantees of the proposed method do not require a specific type of initial solutions.}

The rest of this paper is organized as follows. Section \ref{section 2} first revisits the irrepresentable condition and then proposes the robust regression with coefficient thresholding and nonconvex estimation. Section \ref{section 3} studies theoretical properties of the proposed method, including both statistical guarantees and computational guarantees. The statistical guarantees include the landscape analysis of the nonconvex loss function and asymptotic properties of the proposed estimators, while the computational guarantees concern the convergence of the proposed algorithm. Simulation studies are presented in Section \ref{section 4} and the real application is demonstrated in Section \ref{section 5}. Section \ref{section 6} includes a few concluding remarks.
All the remaining technical details are given in the supplementary file.

\section{Methodology}\label{section 2}
We begin with revisiting the irrepresentable condition and its relaxations in Subsection 2.1. After that, Subsection 2.2 proposes a new high-dimensional robust regression with coefficient thresholding.

\subsection{The Irrepresentable Condition and its Relaxations}
In the high-dimensional linear regression 
$
\bby=\bbX\sbeta+\boldsymbol{\varepsilon},
$
$\bby=(y_1,y_2,\dots, y_n)^T$ is an $n$-dimensional response vector, $\bbX=(\bbx_1,\dots,\bbx_p)$ is an $n\times p$ deterministic design matrix, $\sbeta$ is a $p$-dimensional true regression coefficient vector and  $\boldsymbol{\varepsilon}$ is an error vector with mean zero. We wish to recover the true sparse vector $\sbeta$, whose support $S(\sbeta)
= \{j: \beta_j^{\star}\neq 0\}$ satisfies that its cardinality  $|S(\sbeta)|\ll p$. We use $S$ as the shorthand for the support set $S(\bbeta^{\star})$.

The irrepresentable condition is known to be sufficient and almost necessary for the variable selection consistency of the $\ell_1$ penalization \citep{zhao2006model,zou2006adaptive,wainwright2009sharp}. Specifically, the design matrix $\bbX$ should satisfy that
$$
\underset{ j\in S^c}{\max} \ |\bbx_{j}^T\bbX_S(\bbX_S^T\bbX_S)^{-1}\mathrm{sgn}(\bbeta_S^{\star})| \leq (1-\gamma),
$$
for some incoherence constant $\gamma \in [0,1)$, This condition requires that the variables in the true support are weakly correlated with other variables that are not in the true support. 

There are several versions of relaxation of the irrepresentable condition in the current literature.  \cite{zou2006adaptive} used the adaptive weights $\hat{w}_j$'s in the $\ell_1$ penalization and relaxed the irrepresentable condition as:
$$
\underset{ j\in S^c}{\max} \ \frac{1}{\hat{w}_j}|\bbx_{j}^T\bbX_S(\bbX_S^T\bbX_S)^{-1}\mathbf{\hat{w}}_S\circ\mathrm{sgn}(\bbeta_S^{\star})| \leq (1-\gamma),
$$
where $\circ$ denotes the Hadamard (or componentwise) product of two vectors. The folded concave penalization ~\citep{fan2001variable,zhang2010nearly} also relaxed the irrepresentable condition. Let $p_{\lambda}(\cdot)$ be the folded concave penalty function. Given the local linear approximation and the current solution $\tilde{\bbeta}$, the folded concave penalization relaxed this condition as:
$$
\underset{ j\in S^c}{\max} \ \frac{1}{|\dot{p}_{\lambda}(\tilde{\beta}_j)|}|\bbx_{j}^T\bbX_S(\bbX_S^T\bbX_S)^{-1}|\dot{p}_{\lambda}(\tilde{\bbeta}_S)|\circ\mathrm{sgn}(\bbeta_S^{\star})| \leq (1-\gamma),
$$
where $\dot{p}_{\lambda}(\cdot)$ is the subgradient of the function $p_{\lambda}(\cdot)$. Both the adaptive $\ell_1$ penalization and folded concave penalization utilized the differential penalty functions to relax the restrictive irrepresentable condition. The procedure of adaptive lasso and solving folded concave penalized problem using local linear approximation (LLA) both require a good initialization. Their dependence on the initial solution is inevitably affected by the irrepresentable condition. As a promising alternative, we consider the assignment of adaptive weights on the design matrix $\bbX$, which down-weights the unimportant variables. 
Consider the ideal adaptive weight function $g(\bbeta):=(g(\beta_1),\dots,g(\beta_p))$, where $g(\beta_j)$ goes to $1$ when $j \in S$ and goes to $0$, otherwise. We propose the following nonconvex estimation procedure:
$$
\min_{\bbeta} \frac{1}{2n}\sum_{i=1}^n \{y_i-\sum_{j=1}^p x_{ij}g(\beta_j)\beta_j\}^2+\lambda\|\bbeta\|_1.
$$
Let $\bbz_j=g(\bbeta)\circ \bbx_j$. The irrepresentable condition is now relaxed in this paper as follows:
$$
\underset{ j\in S^c}{\max} \ |g(\beta_{j})\cdot\bbx_{j}^T\bbZ_S(\bbZ_S^T\bbZ_S)^{-1}\circ\mathrm{sgn}(\bbeta_S^{\star})| \leq (1-\gamma).
$$
Given the ideal adaptive weight function $g(\bbeta)$, the proposed approach further relaxes the irrepresentable condition  in comparison to those considered by  existing methods. Unlike existing methods, we propose an integrated nonconvex estimation procedure without depending on  initial solutions.

\subsection{The Proposed Method}

\begin{figure}[h] 
	\centering
	\subfigure[$\tau=0.001$]{\includegraphics[width=1.8in]{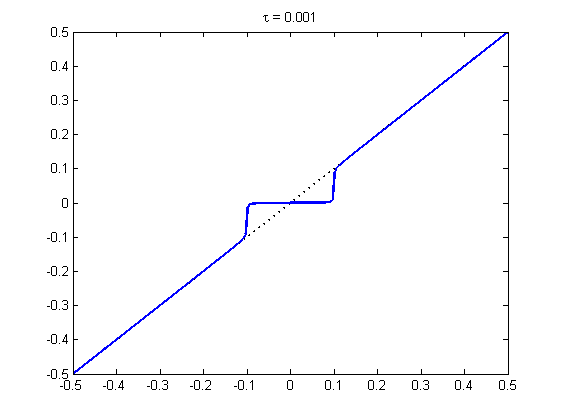} }
	\subfigure[$\tau=0.005$]{\includegraphics[width=1.8in]{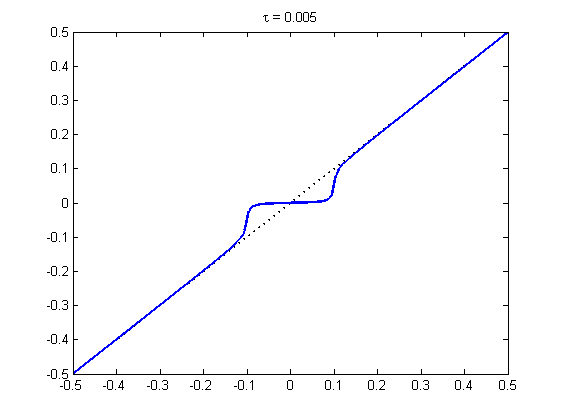} }
	\subfigure[$\tau=0.01$]{\includegraphics[width=1.8in]{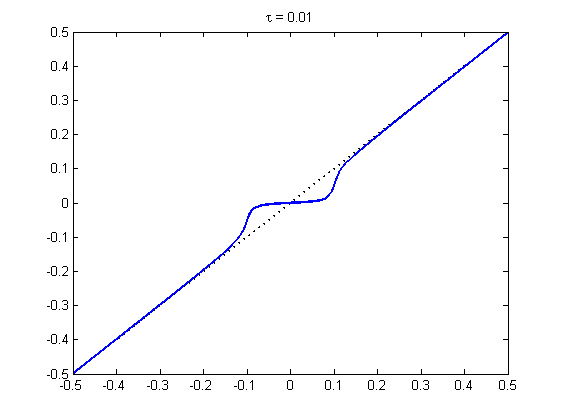} }
	\caption{The smooth approximation of $\beta\circ g_{\tau,\eta}(\beta)$ to $\beta \circ I\{|\beta|\ge \eta\}$ with $\eta=0.1$.}\label{fig:1}
\end{figure}

The weight function plays an important role in relaxing the irrepresentable condition. If we knew the oracle threshold $\eta^{\star}=\underset{j \in S}{\min}\{|\beta_j^{\star}|\}$, the hard thresholding function $I\{|\beta_j|\ge \eta^{\star}\}\}$ would be the ideal choice for the weight function $g(\bbeta)$. However, the discontinuity of 
$I\{|u|\ge \eta\}$ is  challenging for associated optimization. To overcome, we consider a smooth approximation of $I\{|u|\ge \eta\}$ given by
$$
g_{\tau,\eta}(u) = h_\tau(u-\eta)+h_\tau(-u-\eta),
$$
where $h_\tau(w) =1/2+\arctan(w/\tau)/\pi$. Since $h_\tau(w)\to I\{w\ge 0\}$ as $\tau\to 0$ when $w\neq 0$, 
we have $g_{\tau,\eta}(u)\to I\{|u|\ge \eta\}$ as $\tau\to 0$ when $u\neq 0$. Figure \ref{fig:1} illustrates the smooth approximation of $\beta\circ g_{\tau,\eta}(\beta)$ to $\beta \circ I\{|\beta|\ge \eta\}$  when $\tau$ is small (e.g., $\tau=0.001,0.005,0.01).$

Given the above smooth thresholding function $g_{\tau,\eta}(\cdot)$, we propose the robust regression with coefficient thresholding:
\begin{align}
\min_{\bbeta} \frac{1}{n} \sum_{i=1}^n L(y_i-\sum_{j=1}^p x_{ij} \beta_jg_{\tau,\eta}(\beta_j)) \quad \text{subject to} \quad \|\bbeta\|_2\le r,
\end{align}
where following \cite{mei2016landscape} and \cite{loh2017support} we assume that the regression coefficients $\bbeta$ are bounded in the Euclidean ball $\bbB^p(r)\equiv\{\bbeta\in\mR^p: \ \|\bbeta\|_2\le r\}$. Here, to tackle possible outliers, we choose $L(\cdot)$ to be a differentiable and possibly nonconvex robust loss function such as the pseudo Huber loss \citep{charbonnier1997deterministic} or Tukey's biweight loss. For the ease of presentation, throughout this paper, we use the pseudo Huber loss of the following form:
\be\label{huber}
L(a)=\omega^2\{\sqrt{1+(a/\omega)^2}-1\},a\in\mathbb{R},\omega\in\mathbb{R}.
\ee
Note that $L(a)$ provides a smooth approximation of the Huber loss \citep{huber1964robust} and bridges over the $\ell_1$ loss and the $\ell_2$ loss. Specifically, $L(a)$ is approximately an $\ell_2$ loss $a^2/2$ when $a$ is small and an $\ell_1$ loss with slope $\omega$  when $a$ is large. It is worth pointing out that the utility of the pseudo Huber loss essentially adds a weight on each observation $x_{i}$ by $w(x_{i})$ in the following form:
$$
w(x_{i})=\frac{\sqrt{1+(y_i-\hat{y_i})^2/w^2}}{(y_i-\hat{y_i})^2}, 
$$
where $\hat{y_i}$ is a fitted value. Hence, outliers are down-weighted to alleviate potential estimation bias.

To deal with a large number of predictors, we propose the following penalized estimation of high-dimensional robust regression with coefficient thresholding:
\begin{align}\label{prog}
\min_{\bbeta}\left\{ \frac{1}{n} \sum_{i=1}^n L(y_i-\sum_{j=1}^p x_{ij} \beta_jg_{\tau,\eta}(\beta_j))+p_{\lambda}(|\bbeta|)\right\} \quad \text{subject to} \quad \|\bbeta\|_2\le r,
\end{align}
where $p_{\lambda}(|\bbeta|)$ is a chosen penalty function such as the $\ell_1$ penalization or folded concave penalization. In the scalar-on-image analysis, often we need to incorporate some known group information into $p_{\lambda}(|\bbeta|)$. Suppose that the coefficient vector $\bbeta$ is divided into $B$ separate subvectors $\bbeta_1,.\cdots,\bbeta_B$. We can use the group penalty $\lambda\sum_{b=1}^B\|\bbeta_b\|_2$ \citep{yuan2006model} or sparse group penalty $\lambda_1\|\bbeta\|_1+\lambda_2\sum_{b=1}^B\|\bbeta_b\|_2$ \citep{simon2013sparse} in the penalized estimation \eqref{prog}. With $p_{\lambda}(|\bbeta|)=\lambda\sum_{b=1}^B\|\bbeta_b\|_2$, the sparse robust regression with coefficient thresholding and group selection can be written as follows:
\begin{align}\label{prog1}
\min_{\bbeta}\left\{ \frac{1}{n} \sum_{i=1}^n L(y_i-\sum_{j=1}^p x_{ij} \beta_jg_{\tau,\eta}(\beta_j))+\lambda\sum_{b=1}^B\|\bbeta_b\|_2\right\} \quad \text{subject to} \quad \|\bbeta\|_2\le r,
\end{align}
which includes the  $\ell_1$ penalization as a special example with $B=p$. The penalized robust regression with coefficient thresholding \eqref{prog} and \eqref{prog1} can be efficiently solved by a customized composite gradient descent algorithm with provable guarantees, and the details will be presented in Subsection 3.3.

\begin{remark}
Both thresholding function and robust loss work together to relax the restrictive  irrepresentable condition in the presence of possible outliers. Define $\mathbf{W}= ( \mathbf{w}_1,\cdots, \mathbf{w}_p)$ as the weight matrix with $\mathbf{w}_j=(w(x_{1j}),\ldots,w(x_{nj}))$. The irrepresentable condition for the $\ell_1$-penalized robust loss can be written as
$$
\underset{ j\in S^c}{\max} \ | \mathbf{w}_j\circ \bbx_j^T\bbA_S(\bbA_S^T\bbA_S)^{-1}\cdot\mathrm{sgn}(\bbeta_s^{\star})| \leq (1-\gamma),
$$
where $\bbA=\mathbf{W}\circ \bbX$. We impose the weights $\mathbf{W}$ on the row vectors of design matrix $\bbX$. Recall that the thresholding function adds the weights to the column vectors of $\bbX$. Now, with $p_{\lambda}(|\bbeta|)=\lambda\|\bbeta\|_1$ in \eqref{prog}, we add the entry-specific weight $\tilde{\mathbf{W}}=(\tilde{w}_{ij})_{n\times p}$ to $\bbX$ with
$$
\tilde{w}_{ij}=w(x_{ij})\circ g(\beta_j).
$$
Thus, the irrepresentable condition is now relaxed as follows:
$$
\underset{ j\in S^c}{\max} \ |\tilde{\mathbf{w}}_j\circ \bbx_j^T\bbZ_S(\bbZ_S^T\bbZ_S)^{-1}\cdot\mathrm{sgn}(\bbeta_s^{\star})| \leq (1-\gamma),
$$
where $\bbZ=\tilde{\mathbf{W}}\circ \bbX$. By assigning column-wise weight, we allow strong dependence between variables in the true support and other variables; By assigning row-wise weight, we reduce the weights for outliers and enhance the robustness.
\end{remark}

\begin{remark}
The proposed method can be considered as the simultaneous estimation of regression coefficients and adaptive weights to improve the adaptive $\ell_1$ penalization \citep{zou2006adaptive}, whose weights are usually solved from the initial solution or iterated solution. Let $\mathbf{\xi}=\bbeta\circ g_{\tau,\eta}(\bbeta)=(\xi_1,\ldots,\xi_p)^T$. Since $\mathbf{\xi}$ is a continuous and injective function of $\bbeta$ (say, $\mathbf{\xi}=G(\bbeta)$), there is a unique $\bbeta$ such that $\mathbf{\xi}=\bbeta\circ g_{\tau,\eta}(\bbeta)$ holds. Hence, we may write $\bbeta=G^{-1}(\mathbf{\xi})$. The minimization in the proposed high-dimensional robust regression with coefficient thresholding can be rewritten as follows:
\begin{align}\label{prog1b}
\min_{\mathbf{\xi}}\left\{ \frac{1}{n} \sum_{i=1}^n L(y_i-\sum_{j=1}^p x_{ij} \xi_j)+{\lambda}\sum_{j=1}^p\frac{|\xi_j|}{g_{\tau,\eta}(G^{-1}({\mathbf{\xi}})_j)}\right\}. 
\end{align}
However, it does not fall into the folded-concave penalized estimation \citep{fan2001variable,fan2014strong}. The proposed method aims to address the presence of highly correlated predictors, whereas the nonconvex penalized estimation was motivated by correcting the bias of the $\ell_1$-penalized estimation. Also, solving \eqref{prog1b} is extremely challenging, since the denominator $g_{\tau,\eta}(G^{-1}({\mathbf{\xi}})_j)$ can be arbitrarily small when $\xi_j \to 0$. In comparison, the thresholding function is bounded and the non-convex optimization in \eqref{prog1} is computationally tractable.
\end{remark}

\begin{remark} The proposed  method is also related to Bayesian methods~\citep{nakajima2013bayesian_a,nakajima2013bayesian_b,nakajima2017dynamics,kang2018scalar,ni2019bayesian,cai2019bayesian}. In the scalar-on-image regression, \cite{kang2018scalar} proposed a soft-thresholded Gaussian process (STGP) that enjoys posterior consistency for both parameter estimation and variable selection. In this Bayesian framework, $\bbeta$ is assumed to follow an STGP, where the soft thresholding function is defined as 
\begin{equation}
    g_{\eta}(\bbeta)=\mathrm{sgn}(\bbeta)(|\bbeta|-\lambda)\textbf{1}_{\{|\bbeta|>\lambda\}}.
\end{equation}
Thus the regression model can be written as follows:
\begin{align*}
\bby=\bbX \mathrm{sgn}(\bbeta)(|\bbeta|-\lambda)\textbf{1}_{\{|\bbeta|>\lambda\}}(\bbeta)+\boldsymbol{\varepsilon},
\end{align*}
where $\bbeta$ is a realization of a Gaussian process. Compared to \cite{kang2018scalar}, our proposed method is more robust to possible outliers, and uses a very different approach to incorporate the thresholding function that down-weights unimportant variables and achieves sparsity. The proposed method also enjoys both statistical and computational guarantees as detailed in Section 3, whereas the convergence rate of the posterior computation algorithm in  \cite{kang2018scalar} is still unclear. 
\end{remark}

\section{Theoretical Properties}\label{section 3}
This section studies the theoretical properties of our proposed method. After establishing the connection of our proposed method with the thresholded parameter space in Subsection 3.1, we present the landscape analysis and asymptotic properties in Subsection 3.2, and the computational guarantee for an efficient composite gradient descent algorithm in Subsection 3.3.

\subsection{Connection with the Thresholded Parameter Space}

It is known that $\ell_0$-penalization enjoys the oracle risk inequality \citep{barron1999risk}. \cite{fan2013asymptotic2} proved that the global solution of $\ell_0$-penalization, denoted by $\hobeta=(\hat{\beta}^{o}_1,\cdots,\hat{\beta}^{o}_p)^T$, satisfies the hard thresholding property that $\hat{\beta}^{o}_j$ is either $0$ or larger than a positive threshold whose value depends on $\bbX$ and $\bby$ and tuning parameter $\lambda$. Specifically, \cite{fan2013asymptotic2} studied 
the oracle properties of regularization methods over the thresholded parameter space $\mathcal{B_{\eta}}$ defined as
\begin{equation}
    \mathcal{B_{\eta}}=\{\bbeta\in \mathbb{R}^p: \ \|\bbeta\|_2<r,  \|\bbeta\|_0 \le c n/\log p, \text{and for each}\,j, \ \beta_j=0\, \  \text{or}\, \ |\beta_j|\ge \eta\}.
\end{equation}

Given the thresholded parameter space $\mathcal{B_{\eta}}$, we introduce the high-dimensional regression with coefficient hard-thresholding as follows:
	\begin{equation}\label{hard-threg}
\min_{\bbeta} \left\{\frac{1}{n} \sum_{i=1}^n L(y_i-\sum_{j=1}^p x_{ij} \beta_jI\{|\beta_j|\ge \eta\})+p_{\lambda}(|\bbeta|) \right\}\quad \text{subject to} \quad \|\bbeta\|_2\le r.
\end{equation}

This thresholded parameter space guarantees that the hard thresholding property is satisfied with proper $\ell_2$ and $\ell_0$ norm upper bounds. Suppose that global solutions to both (\ref{prog}) and \eqref{hard-threg} exist. It is intuitive that if $\tau$ is very small, an optimal solution to (\ref{prog}) should be ``close" to that of \eqref{hard-threg}. Apparently, $g_{\tau,\eta}(\cdot)$ converges to $I(|\cdot|\ge \eta)$ pointwisely almost everywhere as $\tau \mapsto 0^+$. However it is known that almost surely pointwise convergence does not guarantee the convergence of minimizers (e.g., \citep[Section 7.A]{rockafellar2009variational}). Here we provide the following proposition to show that the global solution of the proposed method~\eqref{prog} converges to the minimizer of coefficient hard-thresholding in \eqref{hard-threg} with additional assumptions.
\begin{prop}\label{prop_1}
	Suppose that $\{\tau^k\}$ is a sequence of scale parameters in the weight function $g_{\tau,\eta}(\cdot)$ such that $\tau^k$ goes to $0$ as $k \to \infty$. Let $\hbeta^k$ be a global minimizer of  \eqref{prog}
	with $\tau=\tau^k$. If 
	\begin{itemize}
	    \item[(i)] $\eta<\min|\beta_j^{\star}|$ and $\|\bbeta^{\star}\|_2\le r$
	    \item[(ii)] For any $\epsilon>0$, there exists $\delta>0$, such that 
	\begin{align*}
	  \max_{1\le j\le p}\mP\left(||\hat{\beta}^k_j|-\eta|<\delta\right)<{\epsilon},
	\end{align*}
	\end{itemize}
    then with arbitrary high probability, the sequence $\{\hbeta^k\}$ enjoys a cluster point $\hbeta$, which is the global minimizer of the high-dimensional regression with coefficient hard-thresholding in \eqref{hard-threg}.
\end{prop}

\begin{remark}
Conditions (i) and (ii) of Proposition \ref{prop_1} are mild in our setting. Condition (i) assumes that the threshold level is smaller than the minimal signal level and the $\ell_2$ norm of true regression coefficients is bounded. Condition (ii) assumes the magnitude of the estimated $\hat{\beta}^k$ is bounded away from the threshold level $\eta$ with high probability, that is, $P\left(||\hat{\beta}^k_j|-\eta|\ge\delta\right)\ge 1-\epsilon$, $\forall j$.  Note that the true regression coefficients $\beta_j^{\star}$'s are either zero or larger than the threshold level $\eta$. As long as the estimator is consistent (see Theorem \ref{theorem5}), Condition (ii) can be satisfied. 
\end{remark}
Given Proposition \ref{prop_1}, we can prove that the total number of falsely discovered signs of the cluster point $\hbeta$, i.e., the cardinality $\mathrm{card}(\{j:\mathrm{sgn}(\hbeta)\neq \mathrm{sgn}(\bbeta^{\star})\})$, converges to zero for the folded concave penalization under the squared loss and sub-Gaussian tail conditions. Recall that the folded concave penalty function $p_{\lambda}(t)$ satisfies that (i) it is increasing and concave in $t \in [0,\infty)$, (ii) $p_{\lambda}(0)=0$, and (iii) it is differentiable with  $\dot{p}_{\lambda}(0_{+})=a\lambda$ for some $a>0$. Define $\kappa_c$ as the smallest possible positive integer such that there exists an $n\times \kappa_c$ submatrix of $n^{-1/2}\bbX$ having a singular value less than a given positive constant $c$. $\kappa_c$ is named robust spark in \citep{fan2013asymptotic2} to ensure model identifiability. Note that $\mathrm{card}(\{j:\mathrm{sgn}(\hbeta)\neq \mathrm{sgn}(\bbeta^{\star})\})$ is an upper bound on the total number of false positives and false negatives. The following proposition implies the variable selection consistency of the cluster point $\hbeta$.

\begin{prop}
	Suppose we re-scale each column vector of the design matrix $\bbX$ for each predictor to have $\ell_2$-norm $n^{1/2}$, and the error $\boldsymbol{\varepsilon}$ satisfies the sub-Gaussian tail condition with mean 0. $\log p=O(n^{c_0})$ and $s=o(n^{1-{c_0}})$ for some constant $c_0\in(0,1)$. And there exists constant $c$ such that $s>\kappa_c/2$.  Let $\mathcal{A}$ be the support of $\bbeta^\star$ and minimal signal strength $\min_{j\in \mathcal{A}}|\beta_j^{\star}| > \eta$. 
	Define $\hbeta$ as the global minimizer of the following optimization problem:
	\begin{align}\label{prog2}
\min_{\bbeta} \left\{\frac{1}{n} \sum_{i=1}^n (y_i-\sum_{j=1}^p x_{ij} \beta_j I\{|\beta_j|\ge \eta\})^2+p_{\lambda}(|\bbeta|)\right\} \quad \text{subject to} \quad \|\bbeta\|_2\le r, \|\bbeta\|_0 \le \kappa_c/2.
\end{align}
If the penalization parameter $\lambda$ is chosen such that $\lambda=c_1\sqrt{\log p/n}=o(\eta)$ for some constant $c_1$, then with probability $1-O(p^{-c_2})$, we have
	$$
	\mathrm{card}(\{j:\mathrm{sgn}(\hbeta)\neq \mathrm{sgn}(\bbeta^{\star})\})\leq Cs\lambda^2\eta^{-2}/(1-C\lambda^2\eta^{-2}),
	$$
where $c_2$ and $C$ are constants.
\end{prop}

Propositions 1 and 2 indicate that the cluster point $\hbeta$ enjoys the similar properties as those of the global solution given by  the $\ell_0$-penalized methods. The proposed method mimics the $\ell_0$-penalization to remove irrelevant predictors and improve the estimation accuracy as $\tau \to 0$. 

{In the sequel, we will show that with high probability our proposed method has a unique global minimizer~(See Theorem 2(b)). 
In this aspect, our proposed regression with coefficient thresholding estimator provides a practical way to approximate the global minimizer over the thresholded space.}

\subsection{Statistical Guarantee}

 After presenting technical conditions, we first provide the landscape analysis and establish the uniform convergence from empirical risk to population risk in Subsection 3.2.1 and then prove the oracle inequalities for the unique minimizer of the proposed method in Subsection 3.2.2.

Now we introduce the notation to simplify our analysis. Let $g(u)$ be a shorthand of $g_{\tau}(u,\eta)$, and let $\gbeta=(\beta_1g(\beta_1),\beta_2g(\beta_2),\cdots,\beta_pg(\beta_p))^{T}$ on $\mR^p$.  Given fixed $\tau$, we claim that $\gbeta$ is third continuously differentiable on its domain. After the direct calculations given in Lemma 1 of the supplementary file, we have the explicit upper and lower bounds for the derivatives of $\gbeta$, i.e.,
\begin{align*}
    &\underline{k_0}I_{p\times p}\preccurlyeq\jbeta\preccurlyeq\overline{k_0}I_{p\times p},\quad\\
    &\Tbeta\preccurlyeq\overline{m_0}I_{p\times p\times p},\quad\\
    &\dbeta\preccurlyeq\overline{s_0}I_{p\times p\times p\times p},
\end{align*}
where $\jbeta:=\nabla\gbeta$, $\Tbeta:=\nabla\jbeta$, and $\dbeta:=\nabla\Tbeta$. Here, $\underline{k_0},\overline{k_0},\overline{m_0}$ and $\overline{s_0}$ are uniform constants independent of $\bbeta$. And $A \preccurlyeq B$ represents that $B-A$ is semi positive definite.

For the group lasso penalty, denote by $\bbeta^a=(\bbeta_1^{a^T},...,\bbeta_{B_1}^{a^T})^T$ the non-zero groups and by $\bbeta^c=(\bbeta_1^{c^T},...,\bbeta_{B_2}^{c^T})^T$ the zero groups, respectively. Following the structure of classical group lasso, let $d_i^a$ and $d_i^c$ be the corresponding length of vector $\bbeta_i^a$ and $\bbeta_i^c$, respectively. Let $d^a_{max}=\underset{i \in \{1,\dots,B_1\}}{\max} \ d_i^a$, $d^c_{max}=\underset{i \in \{1,\dots,B_2\}}{\max} \ d_i^c$, and $d_{\max}=d^a_{\max} \vee d^c_{\max}$, where $d_{\max}$ is finite, not diverging with $n$ and $p$.

We make the following assumptions on the distribution of predictor vector $\bbx$, the true parameter vector $\bbeta^{\star}$ and the distribution of random error  $\boldsymbol{\varepsilon}$.
\begin{assumption} \label{assumption 1}
\begin{itemize}
	\item[(a)] 
	The predictor vector $\bbx$ is $\sigma^2$-sub-Gaussian with mean zero and continuous density $p(\cdot)$ in $\mR^p$, namely $\mE[\bbx]=\mathbf{0}$ and $\mE\{[\exp(\langle\bbu,\bbx\rangle)]\}\le \exp({\sigma^2\|\bbu\|^2_2}/{2})$ for all $\bbu\in\mR^p$. 
	\item[(b)] The feature vector $\bbx$ spans all directions in $\mR^p$, namely $\mE[\bbx\bbx^T]\succcurlyeq\gamma\sigma^2I_{p\times p}$  for some $0<\gamma<1$.
	\item[(c)] 
	The true coefficient vector $\sbeta$ is sparse such that $s_0=\mathrm{supp}(\sbeta)=o(n)$ and  $\|\sbeta\|_2\le r$. Also, the threshold index $\eta$ in weight function $g_{\tau,\eta}(\bbeta)$ satisfies that $\eta\le \min_{j\in S_0}\{|\beta^{\star}_j|\}$.
	\item [(d)] 
	The random error $\boldsymbol{\varepsilon}$ has a symmetric distribution whose density is strictly positive and decreasing on $(0,\infty)$.
\end{itemize}
\end{assumption}
Assumption 1(a) presents the technical conditions on the design matrix. The same conditions were considered in \cite{mei2016landscape} for binary linear classification and robust regression. The sub-Gaussian assumption is a commonly used mild condition in high-dimensional regression. 
Assumption 1(b) imposes the sparsity on the true parameter vector $\sbeta$. Assumption 1(c) permits the possibly heavy tails of the error. We allow the size of the true support set to diverge at rate $o(n)$. Given the sparsity, it is reasonable to limit our theoretical analysis in the Euclidean ball $\Theta_{n,p}=\bbB^p(r)\equiv\{\bbeta\in\mR^p,\|\bbeta\|_2\le r\}$, which can avoid unnecessary technical complications. Also, $\eta$ does not need to be a constant bounded away from $0$ in the asymptotic analysis. Since $G(\bbeta)$ is continuous and injective, there exists a unique $\tsbeta$, such that $\|\tsbeta\|_2\le r/3$ and $\gtbeta=\sbeta$, i.e. $\bbeta^{\star}$ is the thresholded version of $\tilde{\bbeta}^{\star}$. In the sequel, we will study the statistical convergence to the surrogate $\tilde{\bbeta}^{\star}$, which shares the same non-zero support with $\bbeta^{\star}$. In Theorem 2, we show that $\|\hbeta-\tsbeta\|_2=O_p(\sqrt{\frac{s_0\log p}{n}})$. Then $\|G(\hbeta)-G(\tsbeta)\|_2=\|G(\hbeta)-\sbeta\|_2=O_p(\sqrt{\frac{s_0\log p}{n}})$ given $G(\cdot)$ is Lipschitz with fixed $\eta$. In this way, we can use $G(\hbeta)$ as an estimator for $\sbeta$. 
Assumption 1(d) allows random error with heavier tails than the standard Gaussian distribution, and it suits for many applications with outliers. For example, in our simulation, our noise is a mixture distribution with a small variance Gaussian distribution and a large variance Gaussian distribution.  Define $\varphi(\cdot)=L^{\prime}(\cdot)$ and $h(\cdot)=\mE_{\boldsymbol{\varepsilon}}[\varphi(\cdot+\boldsymbol{\varepsilon})]$. Assumption 1(d) can be relaxed as Assumption 1(d') $h(0)=\mE_{\boldsymbol{\varepsilon}}[\varphi(\boldsymbol{\varepsilon})]\ge 0$, which holds for the right skewed error or discrete error. Our theoretical results given in this paper still hold under Assumption 1(a)--(c) and 1(d'). See Remark 6 for more details.

\subsubsection{The Landscape of Population Risk and Empirical Risk}

We analyze the landscape of the population and empirical risk functions of the proposed methods. 
We can bound both gradient vector and Hessian matrix of the population risk $$R(\bbeta)=\mathbb{E}\{L(y-\sum_{j=1}^p x_{j} \beta_jg_{\tau,\eta}(\beta_j))\}$$ and then prove the uniqueness of its stationary point in the following lemma. We let $\nabla R(\bbeta)$ be the gradient of $R(\bbeta)$ with respect to $\bbeta$ and $\nabla^2 R(\bbeta)$ the Hessian matrix. Recall that $\bbB^p(r)=\{\bbeta\in\mR^p,\|\bbeta\|_2\le r\}$. Let
$\bbB^p(\bbeta,\varepsilon_0)$ be the $\ell_2$ ball with the center $\bbeta$ and radius $\varepsilon_0>0$.

\begin{lemma}\label{theorem 1}(Landscape of population risk) Under Assumption \ref{assumption 1}, the population risk function $\prk$ has the following properties:
\begin{itemize}
	\item [(a)] 
	There exists an $\varepsilon_0>0$ and constants $0<\underline{L}_0<\overline{L}_0<\infty$, $T_0>0$ such that
		       \be \label{3}
	\inf_{\bbeta\in\ball\backslash\bbB^p(\tsbeta,\varepsilon_0)} \|\nabla\prk\|_2\ge \underline{L_0},\quad \sup_{\bbeta\in\ball}\|\nabla\prk\|_2\le \overline{L_0};
	\ee
	and for all $\|\bbeta\|_2\le r$,
	\be
	\la\bbeta-\tsbeta,\nabla\prk\ra\ge T_0\|\bbeta-\tsbeta\|_2^2.
	\ee
	\item [(b)]
	For same $\varepsilon_0$ in part (a), there exist constants $0<\underline{M}_0<\overline{M}_0<\infty$ such that
	       \be \label{4}
	       \inf_{\bbeta\in\bbB^p(\tsbeta,\varepsilon_0)} \lambda_{\min}(\nabla^2\prk)\ge \underline{M_0},\quad \sup_{\bbeta\in\ball}\lambda_{\max}(\nabla^2\prk)\le \overline{M_0}.
	      \ee
	      	\item [(c)] 
	      	The population risk $\prk$ has a unique stationary point  $\tsbeta$.
\end{itemize}
\end{lemma}

\begin{remark}
Lemma 1 analyze the landscape of population risk and establishes good properties of population risk function. Indeed, we first show that out of the ball $\sball$, gradient has a lower bound and thus no stationary point exists. In fact, we get a stronger result that no stationary point exists except $\sbeta$. Secondly, inside the ball $\sball$, Hessian matrix of risk function has a lower bound, meaning that $\sbeta$ is a unique minimizer.
\end{remark}

\begin{remark} Lemma 1 still holds under Assumption 1(a)--(c) and 1(d'). In order to establish a lower bound for the gradient of population risk, we need conditions that $h(z)>0$ for all $z>0$ and $h^{\prime}(0)>0$. When we consider the pseudo Huber loss, $\varphi(\cdot)$ is odd and $\varphi(z) > 0$ for all $z>0$; $\varphi^{\prime}(\cdot)$ is even and $\varphi^{\prime}> 0$. If the exchange of expectation and limit are allowed, $h^{\prime}(0)=\mE_{\boldsymbol{\varepsilon}}[\varphi^{\prime}(\boldsymbol{\varepsilon})]>0$ is always true. Thus, Lemma 1 and all the following theorems hold with the same proofs.
\end{remark}

Next we follow  \cite{mei2016landscape} to develop the  uniform convergence results for gradient $\nabla R_n(\bbeta)$ and Hessian $\nabla^2 R_n(\bbeta)$ of empirical risk, to study the landscape of the empirical risk function 
$$
\erk=\frac{1}{n} \sum_{i=1}^n L(y_i-\sum_{j=1}^p x_{ij} \beta_jg_{\tau,\eta}(\beta_j)).
$$
\begin{lemma}\label{theorem 2}(Landscape of empirical risk)
Under Assumption \ref{assumption 1}, for any arbitrarily small $\delta$, there exists a constant $C_1>0$ depending on parameters $(r,\sigma^2,\gamma,\delta,\tau,\eta)$, independent of $n$ and $p$, such that, if $n\ge C_1p\,\text{log}\,p$, the following properties hold for $\erk$, with probability at least $1-\delta$:
\begin{itemize}
	\item[(a)] 
	There exists an $\varepsilon_0>0$ and constants $0<\underline{L}_0<\overline{L}_0<\infty$, $T_0>0$ such that
	\be \label{5}
	\inf_{\bbeta\in\bbB^p(0,r)\backslash\sball} \|\nabla\erk\|_2\ge \underline{L_0}/2,\quad \sup_{\bbeta\in\bbB^p(0,r)}\|\nabla\erk\|_2\le \overline{L_0}/2;
	\ee
	and for all $\bbeta$ satisfying $\|\bbeta-\tsbeta\|_2\ge\varepsilon _0/2$ and $\|\bbeta\|_2\le r$,
	\be
	\la\bbeta-\tsbeta,\nabla\erk\ra\ge \varepsilon_0T_0\|\bbeta-\tsbeta\|_2/4.
	\ee
	\item [(b)]
	For the same $\varepsilon_0$ in part (a), there exist constants $0<\underline{M}_0<\overline{M}_0<\infty$ such that
	       \be \label{6}
	\inf_{\bbeta\in\sball} \lambda_{\min}(\nabla^2\erk)\ge \underline{M_0}/2,\quad \sup_{\bbeta\in\ball}\lambda_{\max}(\nabla^2\erk)\le \overline{M_0}/2,
	\ee
	\item [(c)]
	The empirical risk function $\erk$ has a unique minimizer $\hbeta_n$ satisfying that $$\|\hbeta_n-\tsbeta\|_2\le C_1\sqrt{p\,\log\,n/n}.$$
\end{itemize}
\end{lemma}

It is worth pointing out that Part (c) of Lemma \ref{theorem 2} implies the consistency of the unique minimizer of the empiricial risk function when the dimension $p$ diverges with the sample size $n$ but $n\ge C_1p\,\text{log}\,p$. In the sequel, we will establish the oracle inequality for the proposed penalized robust regression with coefficient thresholding  when the dimension can be much larger than the sample size $n$.

\subsubsection{Oracle Inequality}

Given the landscape analysis, we establish the oracle inequality under the ultra-high dimensional setting when $p$ is at the nearly exponential order of $n$. Specifically, Theorem \ref{theorem4} shows that the sample directional gradient and restricted Hessian converges uniformly to their population counterparts.

 \begin{theorem}\label{theorem4}
 Under Assumption \ref{assumption 1}, for any arbitrarily small $\delta$, there exist constants $C_2$ and $C_3$ that depends on $(r,\sigma^2,\gamma,\delta,\tau,\eta)$ such that the following properties hold:
 \begin{itemize}
 	\item[(a)] The empirical directional gradient converges uniformly to population directional gradient along the direction $(\bbeta-\tsbeta)$, namely
 	\be
 	\mP\left(\sup_{\bbeta\in\ball/\{\mathbf{0}\}}\frac{|\la\nabla\erk-\nabla\prk,\bbeta-\tsbeta\ra|}{\|\bbeta-\tsbeta\|_1}\le\sigma\sqrt{\frac{C_2\log(p)}{n}}\right)\ge 1-\delta.
 	\ee
 	\item[(b)] The empirical restricted Hessian converges uniformly to the population restricted Hessian in the set $\ball \cap \bbB^p_0(s_0)$ for any $s_0\le p$. That is, if $\|\bbeta\|_2<r$ and $\|\bbeta\|_0<s_0$, as $n\ge C_3s_0\log p$, we have
 	\be
 	\mP\left(\sup_{(\bbeta,\bbv)\in\Omega}|\la\bbv,(\nabla^2\erk-\nabla^2\prk)\bbv\ra|\le\sigma^2\sqrt{\frac{C_3s_0\log(p)}{n}} \right)\ge 1-\delta,
 	\ee
 	 	where $\Omega_1=\ball\cap\oball$, $\Omega_2=\bbB^p(1)\cap\oball$ and $\Omega=\Omega_1\times\Omega_2$.
 \end{itemize}
  \end{theorem}

 Theorem \ref{theorem5} proves the oracle inequality of the global minimizer of the proposed high-dimensional robust regression with coefficient thresholding as well as its uniqueness with high probability.
 \begin{theorem}\label{theorem5}
 	Under Assumption \ref{assumption 1}, for any arbitrarily small $\delta$, there exist constants $C_4$, $C_5$ and $C_{\lambda}$ that depends on 
 	$(r,\sigma^2,\gamma,\eta,\delta,\tau,M,t,d_{\max}^a,d_{\max}^c)$ but are independent of $n,p,s_0$, such that as $n\ge C_5s_0\log p$ and $\lambda\ge C_{\lambda}\sqrt{(\log p)/n}$, the following properties hold with probability at least $1-\delta$:
 	\begin{itemize}
 		\item  [(a)] Any stationary point $\hbeta$ of group-regularized risk minimization satisfies
 		\be
 		\|\hbeta-\tsbeta\|_2\le C_4\sqrt{(s_0\log p)/n+s_0\lambda^2}.
 		\ee
 	    \item [(b)] Furthermore, let $r_s=C_4\sqrt{(s_0\log p)/n+s_0\lambda^2}$.  There exists a constant $C_6$, for $n\ge C_6\log^2p$ such that  $r_s \le \varepsilon_0$, the optimization problem \eqref{prog1} has a unique global minimizer $\hbeta$.
      \end{itemize}
 \end{theorem}

\subsection{Computational Guarantee}

Gradient descent algorithms do not work since the objective function is not differentiable at zero. We consider the composite gradient descent algorithm \citep{nesterov2007gradient}, which is computationally efficient for solving the non-smooth and nonconvex optimization and enjoys the desired convergence property. Specifically, solving the proposed robust regression with coefficient thresholding on proposed composite gradient descent algorithm consists of two key steps at each iteration: the gradient descent step and the $\ell_2$-ball projection step. To perform the step of gradient descent, we derive the gradient vector of the empirical risk function:
$$
\nabla\widehat{R}_n(\bbeta)=\sum_{i=1}^n L^{'}(\gbeta^T\bbx_i-y_i)\bbx_i^T\jbeta,
$$
where $\jbeta=\text{diag}(g(\beta_1)+\beta_jg'(\beta_1),\cdots,g(\beta_p)+\beta_jg'(\beta_p))$ is a diagonal matrix. Given the previous iterated solution $\hbeta^{(k)}$, we need to solve the following subproblem:
$$
\min_{\bbeta} \ \left \{\frac{1}{2}\|\bbeta-(\hbeta^{(k)}-\frac{1}{h}\nabla \hat{R}(\hbeta^{(k)}))\|^2_2+\frac{\lambda}{\eta}\sum_{b=1}^B\|\bbeta_b\|_2 \right \}.
$$
It is known that this subproblem has a closed-form solution through the following group-wise soft thresholding operator:
$$
S_{t}(\bxi)=\bxi_j\cdot \|\bxi_j\|_2^{-1} \circ (\|\bxi\|_2-t)_{+},
$$
where $\circ$ denotes the Hadamard product. Thus, the gradient descent step can be solved as:	$$\tbeta^{(k+1)}=S_{\lambda/h}(\hbeta^{(k)}-h\nabla(\widehat{R}_n(\hbeta^{(k)}))).$$
Next, in the second step, we perform the following projection onto the $\ell_2$-ball:
$$
\underset{\bbeta}{\min} \
 \|\bbeta-\tilde{\bbeta}^{(k+1)}\|_2 \quad \text{subject to } \ \|\bbeta\|_2 \leq r,
$$
which is given by the following closed-form projection:
$$
\pi_r(\tbeta^{(k+1)})=\frac{\min\{\|\tbeta^{(k+1)}\|_2,r\}}{\|\tbeta^{(k+1)}\|_2}\tbeta^{(k+1)}.
$$
To sum up, the proposed composite gradient descent algorithm can be proceeded as Algorithm 1.

\smallskip

\begin{algorithm}[H]
{\textbf{Input}: $\bbeta^{(0)}\in\ball$, step size $h$, penalization parameter $\lambda$, and thresholding parameter $\eta$\\}
	\For{$k=0, 1, 2, ...$ until convergence}{
		$\tbeta^{(k+1)}=S_{\lambda/h}(\hbeta^{(k)}-h\nabla(\widehat{R}_n(\hbeta^{(k)})))$\\
		$\hbeta^{(k+1)}=\pi_r(\tbeta^{(k+1)})$
	}
	\caption{The proposed composite gradient descent algorithm.}
\end{algorithm}

\smallskip

At each iteration, the subproblem can be solved easily with the closed-form solution, and the computational complexity is on the quadratic order of dimension $p$. The algorithmic convergence rate 
is presented in the following theorem. 

\begin{theorem} 
Let $\hat{\bbeta}^{(k)}$ be the $k$-th iterated solution of Algorithm 1. 
There exist constants $c_r$,$c_h$ and $C_7$ are independent of $(n,p,s_0)$ such that when we choose $r>c_r$ and $h<c_h$, we can find one iteration $k$ that can be upper bounded by $ C_7\frac{1}{\epsilon^2}$, such that there exists a subgradient we denote as $g(\hbeta^{(k)}_b) \in \partial \|\hbeta^{(k)}_b\|_2$:
$$
 \|\nabla \widehat{R}_n(\hbeta^{(k)})+\lambda \sum_{b=1}^Bg(\hbeta^{(k)}_b)\|_2 \leq \epsilon
$$
where $\partial \|\bbeta_b\|_2$ denotes the sub-differential of the group penalty function \citep{rockafellar2015convex}.
\end{theorem}

\begin{remark}
Theorem 3 provides a theoretical justification of the algorithmic convergence rate. More specifically, the proposed algorithm always converges to an approximate stationary solution (a.k.a. $\epsilon$-stationary solution) at finite sample sizes. In other words, for the solution solved after $O(\frac{1}{\epsilon^2})$ iterations, the subgradient of the objective function is bounded by $\epsilon$ under the $\ell_2$ norm  at finite sample sizes. When $k$ increases, the proposed algorithm will find the stationary solution that satisfies the subgradient optimality condition as $\epsilon \to 0$. To better understand the algorithmic convergence in practice, we also provide the convergence plots and computational cost of the proposed algorithm in simulation studies in Section  C of the supplementary file.
Combining Theorems 2 and 3, we may obtain the convergence guarantee  with high probability. From both theoretical and practical aspects, the proposed algorithm is computationally efficient and achieves the desired computational guarantee. Observing that the nice empirical gradient structure proved in Theorem 2, in Appendix Section C, we further provide Theorem 4, which is an extension of Theorem 3. It further shows that given the solution is sparse, the composite gradient descent algorithm guarantees a linear convergence rate of $\bbeta$, with respect to the number of iterations.
\end{remark}

\section{Simulation Studies}\label{section 4}

This section carefully examines the finite-sample performance of our proposed method in simulation studies. To this end, we compare the proposed robust estimator with coefficient thresholding (denoted by RCT) to Lasso, Adaptive Lasso (denoted by AdaLasso), SCAD and MCP penalized estimators in eight different linear regression models (Models 1--8) and with Lasso, Group Lasso (denoted by GLasso), and Sparse Group Lasso (denoted by SGL) in two Gaussian process regression models (Models 9 and 10). Models 9 and 10 mimic the scalar-on-image regression analysis. We implemented the Lasso and the Adaptive lasso estimators using R package `glmnet', and chose the weight for adaptive lasso estimator based on the Lasso estimator. Group lasso is implemented using the method in \citep{yang2015unified}. SCAD and MCP penalized estimators are implemented using R package `ncvreg'. And we also verify that the estimation results are consistency with R package `Picasso'.

We compare estimation performance based on the root-mean-square error (RMSE, that is $\|\hat{\bbeta}-\bbeta^{\star}\|_2$) and variable selection performance based on both false positive rate (FPR) and false negative rate (FNR).  $\text{FPR}={\text{FP}}/({\text{FP}+\text{TN}})$, and $\text{FNR}={\text{FN}}/({\text{FN}+\text{TP}})$, where TN, TP, FP and FN represent the numbers of true negative, true positive, false positive and false negative respectively. Each measure is computed as the average over $50$ independent replications.

Before proceeding to simulation results, we discuss the choice of tuning parameters for the proposed method. The proposed method involves tuning over the  penalization parameter $\lambda$, coefficient thresholding parameter $\eta$, step size $h$, and radius $r$ of the feasible region. Here, $h$ is chosen to be relatively small such that the algorithm does not encounter overflow issues. To simplify the tuning process, we set $h$ to be a small constant in our empirical studies; alternatively,  $h$ can also be chosen according to an acceleration process in \cite{nesterov2007gradient} to further achieve faster convergence. Also, $r$ is chosen to be relatively large to specify the appropriate feasible region, and $\eta$ can be specified as any arbitrary quantity that is smaller than the minimal signal strength. In addition, $\tau$ is chosen to be properly small to make our soft-thresholding function mimic the true thresholding function. It is important to point out that the algorithmic convergence and numerical results are generally robust to the choices of $\eta$, $h$, and $r$. In our simulation studies, we choose $h=0.01$, $r=20$, $\tau=0.01$. With these prespecified $h$, $r$, and $\tau$, we choose the penalization parameter $\lambda$ and $\eta$ using the $5$-folded cross-validation based on $\ell_1$ prediction error. Additionally, $\eta$ can also be chosen as the $30\%$ quantile of the absolute values of non-zero coefficients of Lasso.

\subsection{Linear Regression}
First, we simulated data from the linear model $y=\bbx'\bbeta+\varepsilon$ where $\bbx \sim N(0,\bSigma)$. We consider different correlation  structures for $\bSigma = (\sigma_{ij})_{p\times p}$ in the following six simulation models:
\begin{align*}
&\text{Model 1}: \sigma_{ij}=0.5^{|i-j|},\ \qquad\qquad\text{AR1(0.5)} \\
&\text{Model 2}: \sigma_{ij}=0.6^{|i-j|},\     \qquad\qquad\text{AR1(0.6)}  \\
&\text{Model 3}: \sigma_{ij}=0.7^{|i-j|},\    \qquad\qquad\text{AR1(0.7)} \\
&\text{Model 4}: \sigma_{ij}=0.4+0.6I(i=j),\   \text{CS(0.4)}  \\
&\text{Model 5}: \sigma_{ij}=0.5+0.5I(i=j),\   \text{CS(0.5)}  \\
&\text{Model 6}: \sigma_{ij}=0.6+0.4I(i=j),\   \text{CS(0.6)}.
\end{align*}
Models 1--3 have autoregressive (AR) correlation structures, in which the irrepresentable condition holds for Model 1 but fails for Models 2 and 3. Models 4-6 have the compound symmetry (CS) correlation structures and the irrepresentable condition fails in all these models. 

We assume a mixture model for the error, $\varepsilon \sim 0.9N(0,\sigma_1^2)+0.1N(0,\sigma_2^2)$, where  $\sigma^2_2$ is set much larger than $\sigma^2_1$. For Models 1--3, set $\sigma^2_2=10$ and $\sigma^2_1=1, 2$ or $3$ in  case (a), (b) or (c), respectively.  For Models 4--6, set $\sigma^2_2=3$ and $\sigma^2_1=0.1$,$0.3$ or $1$ in  case (a), (b) or (c), respectively. For all the models, we choose $n=100$, $p=2000$ and $\sbeta=(\textbf{1}_{20},\textbf{0}_{1980})$ to create a high dimensional regime.


\begin{table}[]
\center
\caption{Estimation and selection accuracy of different methods for Models 1--3}
\scalebox{0.9}{
{\footnotesize
\begin{tabular}{cccc|ccc|ccc}
\hline
                                                        &  FPR   &FNR   & $\ell_2$ loss      &  FPR   &FNR   & $\ell_2$ loss  &  FPR   &FNR   & $\ell_2$ loss \\ \hline
                                                                                   & \multicolumn{3}{c|}{Model (1a)} & \multicolumn{3}{c|}{Model (2a)} & \multicolumn{3}{c}{Model (3a)} \\ \hline
\multicolumn{1}{l|}{\multirow{2}{*}{Lasso}} &   0.021     &  0.199    &    3.198      &   0.017     &  0.165   & 3.032  &    0.014&0.153&3.035 \\
\multicolumn{1}{l|}{}                                  &(0.009) &(0.130)& (0.475)  &(0.010) &(0.100)& (0.407)   & (0.008)&(0.091)&(0.391) \\
\multicolumn{1}{l|}{\multirow{2}{*}{AdaLasso}}                        &0.020           &0.212    &  3.787    &  0.016   & 0.180       & 3.716 &0.014 &0.156 &3.739   \\
\multicolumn{1}{l|}{}                                & (0.008)&(0.143)  &(0.821) &(0.007)& (0.111)&(0.816) &(0.008) &(0.111) & (0.815)  \\
\multicolumn{1}{l|}{\multirow{2}{*}{SCAD}}                   & 0.007         & 0.422     &   4.148    & 0.008   &  0.474      & 4.750 & 0.010 & 0.575 & 5.979    \\
\multicolumn{1}{l|}{}                                 & (0.006)&(0.137)  &(0.950) &(0.007)& (0.123)&(1.032)  &(0.006) &(0.107) &(1.054) \\ 
\multicolumn{1}{l|}{\multirow{2}{*}{MCP}}                   & 0.003         & 0.625     &   4.709    &  0.004   &  0.654      & 5.430  & 0.003 & 0.694 & 6.201    \\
\multicolumn{1}{l|}{}                                 & (0.002)&(0.065)  &(0.537) &(0.002)& (0.060)&(0.571)  &(0.002) &(0.049) &(0.536) \\ 
\multicolumn{1}{l|}{\multirow{2}{*}{RCT}}                   & 0.010         & 0.177     &   2.860    &  0.004   &  0.071      & 2.041 & 0.002 & 0.018 & 1.466   \\
\multicolumn{1}{l|}{}                                 & (0.008)&(0.117)  &(0.888) &(0.002)& (0.112)&(0.754)  &(0.002) &(0.035) &(0.502) \\ \hline
          & \multicolumn{3}{c|}{Model (1b)} & \multicolumn{3}{c|}{Model (2b)} & \multicolumn{3}{c}{Model (3b)} \\ \hline
\multicolumn{1}{l|}{\multirow{2}{*}{Lasso}} &   0.019    &  0.243    &    3.446      &   0.017     &  0.209   & 3.319  &    0.014&0.187&3.273 \\
\multicolumn{1}{l|}{}                                  &(0.010) &(0.140)& (0.372)  &(0.008) &(0.101)& (0.406)   & (0.008)&(0.087)&(0.422) \\
\multicolumn{1}{l|}{\multirow{2}{*}{AdaLasso}}                        &0.018           &0.256   &  4.134    &  0.016   & 0.330       & 4.027 &0.013 &0.199 &4.062    \\
\multicolumn{1}{l|}{}                                & (0.008)&(0.130)  &(0.657) &(0.008)& (0.104)&(0.687) &(0.008) &(0.107) & (0.726)  \\
\multicolumn{1}{l|}{\multirow{2}{*}{SCAD}}                   & 0.008         & 0.443     &   4.233    & 0.008   &  0.477      & 4.772 & 0.009 & 0.566 &5.726   \\
\multicolumn{1}{l|}{}                                 & (0.005)&(0.140)  &(0.998) &(0.006)& (0.156)&(1.236)  &(0.007) & (0.153) & (1.393) \\ 
\multicolumn{1}{l|}{\multirow{2}{*}{MCP}}                   & 0.003         & 0.636     &   4.771    &  0.003   &  0.674      & 5.573 & 0.003 & 0.708 & 6.386    \\
\multicolumn{1}{l|}{}                                 & (0.002)&(0.069)  &(0.627) &(0.002)& (0.060)&(0.577)  &(0.002) &(0.058) &(0.667)\\ 
\multicolumn{1}{l|}{\multirow{2}{*}{RCT}}                   & 0.018         & 0.242     &   3.879    &  0.010   &  0.154      & 3.331 & 0.007 & 0.084 & 2.939    \\
\multicolumn{1}{l|}{}                                 & (0.016)&(0.163)  &(0.735) &(0.009)& (0.098)&(0.691)  &(0.005) &(0.127) &(0.755) \\ \hline
          & \multicolumn{3}{c|}{Model (1c)} & \multicolumn{3}{c|}{Model (2c)}  & \multicolumn{3}{c}{Model (3c)}\\ \hline
\multicolumn{1}{l|}{\multirow{2}{*}{Lasso}} &   0.018     &  0.338    &    3.706      &   0.017     &  0.244   & 3.589  &    0.014 &0.242 &3.611 \\
\multicolumn{1}{l|}{}                                  &(0.010) &(0.162)& (0.383)  &(0.010) &(0.106)& (0.475)   & (0.010)&(0.093)&(0.450) \\
\multicolumn{1}{l|}{\multirow{2}{*}{AdaLasso}}                        &0.019            &0.195    &  4.148    &  0.016   & 0.258       & 4.469 &0.012 &0.243 &4.459    \\
\multicolumn{1}{l|}{}                                & (0.018)&(0.156)  &(0.650) &(0.010)& (0.114)&(0.590) &(0.009) &(0.105) & (0.590)  \\
\multicolumn{1}{l|}{\multirow{2}{*}{SCAD}}                   & 0.008         & 0.498     &   4.390    &  0.007   &  0.481      & 4.722 & 0.006& 0.549 & 5.599    \\
\multicolumn{1}{l|}{}                                 & (0.005)&(0.134)  &(1.034) &(0.005)& (0.147)&(1.331)  &(0.006) &(0.149) &(1.612) \\ 
\multicolumn{1}{l|}{\multirow{2}{*}{MCP}}                   & 0.003         & 0.678     &   4.968    &  0.003   &  0.697      & 5.624 & 0.003& 0.732 & 6.506    \\
\multicolumn{1}{l|}{}                                 & (0.001)&(0.069)  &(0.716) &(0.002)& (0.054)&(0.709)  &(0.002) &(0.047) &(0.823) \\ 
\multicolumn{1}{l|}{\multirow{2}{*}{RCT}}                   & 0.025         & 0.294     &   4.305    &  0.019   &  0.195     & 4.148 & 0.011 & 0.164 & 3.886    \\
\multicolumn{1}{l|}{}                                 & (0.156)&(0.230)  &(0.574) &(0.018)& (0.156)&(0.650)  &(0.011) &(0.136) &(0.688) \\ \hline
\hline
\end{tabular}
}
}
\end{table}

\begin{table}[]
\center
\caption{Estimation and selection accuracy of different methods for Models 4--6}
\smallskip
\scalebox{0.9}{
{\footnotesize
\begin{tabular}{cccc|ccc|ccc}
\hline
                                                        &  FPR   &FNR   & $\ell_2$ loss      &  FPR   &FNR   & $\ell_2$ loss  &  FPR   &FNR   & $\ell_2$ loss \\ \hline
                                                                                   & \multicolumn{3}{c|}{Model (4a)} & \multicolumn{3}{c|}{Model (5a)} & \multicolumn{3}{c}{Model (6a)} \\ \hline
\multicolumn{1}{l|}{\multirow{2}{*}{Lasso}} &   0.040     &  0.337    &    4.075      &   0.041     &  0.387   & 4.183  &    0.041 & 0.374 & 4.144 \\
\multicolumn{1}{l|}{}                                  &(0.004) &(0.135)& (0.501)  &(0.003) &(0.126)& (0.458)   & (0.004)&(0.135)&(0.422) \\
\multicolumn{1}{l|}{\multirow{2}{*}{AdaLasso}}                        &0.033           &0.369    &  4.035    &  0.033   & 0.421      & 3.726 &0.032 &0.443 &4.512   \\
\multicolumn{1}{l|}{}                                & (0.003)&(0.138)  &(0.626) &(0.003)& (0.134)&(0.578) &(0.003) &(0.135) & (0.763)  \\
\multicolumn{1}{l|}{\multirow{2}{*}{SCAD}}                   & 0.021        & 0.736     &   5.849     & 0.021        & 0.736     &   5.849   & 0.020 & 0.745 & 5.711    \\
\multicolumn{1}{l|}{}                                 & (0.007)&(0.147)  &(1.238)  & (0.007)&(0.147)  &(1.238)  &(0.009) &(0.149) &(1.151) \\ 
\multicolumn{1}{l|}{\multirow{2}{*}{MCP}}                   & 0.008        & 0.868     &   6.763   & 0.008   & 0.896     &   7.129 & 0.007 & 0.921 & 7.348    \\
\multicolumn{1}{l|}{}                                 & (0.003)&(0.084)  &(0.558)  & (0.003)&(0.065)  &(0.541)  &(0.003) &(0.064) &(0.510) \\ 
\multicolumn{1}{l|}{\multirow{2}{*}{RCT}}                   & 0.061         & 0.215     &   3.982    &  0.062   &  0.244      & 4.023 & 0.066 & 0.253 & 4.093   \\
\multicolumn{1}{l|}{}                                 & (0.007)&(0.089)  &(0.265) &(0.009)& (0.093)&(0.283)  &(0.009) &(0.098) &(0.314) \\ \hline
          & \multicolumn{3}{c|}{Model (4b)} & \multicolumn{3}{c|}{Model (5b)} & \multicolumn{3}{c}{Model (6b)} \\ \hline
\multicolumn{1}{l|}{\multirow{2}{*}{Lasso}} &   0.040    &  0.351   &   4.092      &   0.041     &  0.378   & 4.173  &    0.041& 0.364& 4.124 \\
\multicolumn{1}{l|}{}                                  &(0.004) &(0.135)& (0.448)  &(0.004) &(0.125)& (0.451)   & (0.003)&(0.081)&(0.235) \\
\multicolumn{1}{l|}{\multirow{2}{*}{AdaLasso}}                        &0.033           &0.377   &  4.083    &  0.033   & 0.411       & 4.283 &0.033 &0.459 &4.553    \\
\multicolumn{1}{l|}{}                                & (0.003)&(0.139)  &(0.561) &(0.003)& (0.134)&(0.565) &(0.003) &(0.145) & (0.819)  \\
\multicolumn{1}{l|}{\multirow{2}{*}{SCAD}}                   & 0.022         & 0.719     &   5.563   & 0.021         & 0.722     &   5.668 & 0.020 & 0.744 & 5.802    \\
\multicolumn{1}{l|}{}                                 & (0.022)&(0.135)  &(1.178)  & (0.007)&(0.138)  &(1.154)  &(0.008) &(0.157) &(1.229) \\ 
\multicolumn{1}{l|}{\multirow{2}{*}{MCP}}                   & 0.008        & 0.868     &   6.738    & 0.007        & 0.900     &   7.142  & 0.008 & 0.900 & 7.146    \\
\multicolumn{1}{l|}{}                                 & (0.003)&(0.079)  &(0.600)  & (0.003)&(0.065)  &(0.614)  &(0.004) &(0.080) &(0.719) \\ 
\multicolumn{1}{l|}{\multirow{2}{*}{RCT}}                   & 0.060         & 0.226     &   4.019    &  0.063   &  0.260      & 4.067 & 0.066 & 0.275 & 4.138    \\
\multicolumn{1}{l|}{}                                 & (0.008)&(0.090)  &(0.249) &(0.007)& (0.191)&(0.181)  &(0.007) &(0.089) &(0.255) \\ \hline
          & \multicolumn{3}{c|}{Model (4c)} & \multicolumn{3}{c|}{Model (5c)}  & \multicolumn{3}{c}{Model (6c)}\\ \hline
\multicolumn{1}{l|}{\multirow{2}{*}{Lasso}}    &  0.040    &    0.382      &   4.173     &  0.041   & 0.418  & 4.265 &   0.040 &0.408 &4.324 \\
\multicolumn{1}{l|}{}                                  &(0.003) &(0.131)& (0.385)  &(0.004) &(0.133)& (0.360)   & (0.003)&(0.103)&(0.352) \\
\multicolumn{1}{l|}{\multirow{2}{*}{AdaLasso}}                        &0.034            &0.411    &  4.264    &  0.033   & 0.446       & 4.347 &0.032 &0.478 &4.633    \\
\multicolumn{1}{l|}{}                                & (0.002)&(0.135)  &(0.473) &(0.003)& (0.127)&(0.446) &(0.003) &(0.140) & (0.549)  \\
\multicolumn{1}{l|}{\multirow{2}{*}{SCAD}}                   & 0.023        & 0.704     &   5.343   & 0.021        & 0.726     &   5.688 & 0.021 & 0.727 & 5.612   \\
\multicolumn{1}{l|}{}                                 & (0.006)&(0.148)  &(1.168)   & (0.007)&(0.164)  &(1.415)  &(0.008) &(0.179) &(1.390) \\ 
\multicolumn{1}{l|}{\multirow{2}{*}{MCP}}                   & 0.008        & 0.885     &   6.979    & 0.007        & 0.905     &   7.217 & 0.007 & 0.915 & 7.441    \\
\multicolumn{1}{l|}{}                                 & (0.004)&(0.073)  &(0.531)   & (0.004)&(0.069)  &(0.578)  &(0.027) &(0.062) &(0.411) \\ 
\multicolumn{1}{l|}{\multirow{2}{*}{RCT}}                   & 0.061         & 0.267     &   4.147    &  0.062   &  0.290     & 4.228 & 0.064 & 0.314 & 4.275   \\
\multicolumn{1}{l|}{}                                 & (0.007)&(0.102)  &(0.239) &(0.007)& (0.173)&(0.212)  &(0.008) &(0.110) &(0.228) \\ \hline
\hline
\end{tabular}
}
}
\end{table}

Tables 1 and 2 summarize the simulation results for Models 1--6. Several observations can be made from this table. First, in Models 1--3, as the auto correlation increases, it becomes clearer that our robust  estimator with coefficient thresholding (RCT) outperforms both Lasso ,Adaptive Lasso (AdaLasso) and nonconvex penalized estimators (SCAD and MCP). Nonconvex penalized estimators do not work well on all these settings, since they tend to choose too-large penalization, and it results in a much higher false negative rate than other methods. The Lasso estimator misses many true signals due to the high correlated $\bbx$, leading to that adaptive lasso also cannot perform well given the Lasso initials. Especially in Model 3, when the auto correlation is high, our thresholding estimator has much smaller FPR and FNR compared to lasso-type methods. Second, in case (a), where the effects of outliers are stronger than (b) and (c),  our estimator achieves a relatively better performance  than other two estimators thanks to the use of the pseudo Huber loss. Third, in more challenging Models 4--6,  our estimator is able to identify more true predictors and well controls  false positives. In summary, the proposed estimator is more favorable in high dimensional regression setting, especially when the predictors are highly correlated, such as AR1(0.7) model.

As we introduced in previous sections, nonconvex penalization such as MCP penalized regression does not require irrepresentable condition. However, in our simulation, we found that it does not work well in our setting. We find the major reason is that it does not work not as well as our estimator when dimension is too large comparable the sample size: $n=100$, $p=2000$ in our case. Here we summarize the regression simulation settings for major nonconvex penalized regression papers. \cite{breheny2011coordinate} has $n=100$ and $p=500$ for linear regression setting. It has larger dimension of features under logistic regression setting, but nonconvex penalized estimator performs no better than $\ell_1$ penalized estimator in the aspect of prediction. \cite{zhang2010nearly} has $n=300$ and $p=2000$, with features being generated independently, in which the irrepresentable condition is not violated. \cite{fan2014strong} used a setting of $n=100$ and $p=1000$ with AR1(0.5) covariance matrix, which does not violate the irrepresentable condition. \cite{loh2017support} used $p=512$ and various different sample sizes larger than $100$. \cite{wang2014optimal} used a setting of $n=200$ and $p=2000$. 
In comparison to these reported simulation studies, our setting of $n=100$, $p=2000$ appears to be more challenging, in which larger variances in the generative model add additional difficulty in the analysis. Therefore MCP/SCAD penalized estimators do not perform well in these settings considered in our simulation studies, which clearly demonstrate the advantage of our proposed method.

\subsection{Gaussian Process Regression}

We now report simulation results to mimic the scalar-on-image regression. In the simulation, we consider a two-dimensional image structure and the whole region is generated by Gaussian processes. In these models,  let $n=500$ and $p=50\times 50$, and consider the following covariance structures:

\begin{align*}
&\text{Model 7} \ (\text{GP1(10)}): \quad \sigma_{ij}=\exp(-\|s_i\|^2-\|s_j\|^2-10\|s_i-s_j\|^2),\  \\
&\text{Model 8} \ (\text{GP1(5)}): \quad \sigma_{ij}=\exp(-\|s_i\|^2-\|s_j\|^2-5\|s_i-s_j\|^2), 
\end{align*}
where $s_i$ and $s_j$ are two points in the rectangle $[-1,1]\times[-1,1]$.
And the coefficients with nonzero effects are in a circle of the graph, with radius 0.1 for the true parameters. The values of the nonzero effects are generated by a uniform distribution on $[0.5,1]$, similar to the setting given in \cite{heselective}. The errors follow the same mixture model, $\varepsilon \sim 0.9N(0,\sigma_1^2)+0.1N(0,\sigma_2^2)$. For Models 7--8, set $\sigma^2_2=30$, and refer to case (a), (b) and (c) with $\sigma^2_1 = 2,4$ and $8$, respectively. A typical realization of $\bbx$ for Model 7 is illustrated in Figure 3(a). 

In this section, we also evaluate the soft-threholded Gaussian Process(STGP) estimator\citep{kang2018scalar}, which is a state-of-art scalar-on-image regression method. STGP assumes a spatial smootheness condition, which is satisfied by Model 7-10. The original STGP estimator requires knowledge of whether any two predictors are adjacent. Such information may be unavailable or cannot be implemented in many cases. For example, when the number of voxels are too large to be fully processed, we may have to preselect a subset of voxels or use certain proper dimension reduction techniques to overcome related computational difficulties. Given that all other methods do not need such adjacency information, to fairly compare STGP with other methods, we consider two STGP estimators in our comparison. That is, STGP represents the original STGP estimator; STGP(no-info) represents STGP estimator without using adjacency information.

\begin{table}[]
\center
\caption{Estimation and selection accuracy of different methods for Models 7 and 8}
{\footnotesize
\begin{tabular}{cccc|ccc}
\hline
                                                        &  FPR   &FNR    & $\ell_2$ loss        &  FPR   &FNR    & $\ell_2$ loss    \\ \hline
                                                                                   & \multicolumn{3}{c|}{Model (7a)} & \multicolumn{3}{c}{Model (8a)}   \\ \hline
\multicolumn{1}{l|}{\multirow{2}{*}{Lasso}} &   0.002     &  0.814   &    7.083     &   0.001     &  0.784   & 6.071         \\
\multicolumn{1}{l|}{}                                  &(0.002) & (0.031) &(1.561)   &(0.001) &(0.068)& (1.305)    \\
\multicolumn{1}{l|}{\multirow{2}{*}{AdaLasso}}                        &0.002          & 0.820 &  12.527    &  0.001   & 0.784      & 0.196\\
\multicolumn{1}{l|}{}                                & (0.031)&(0.031)  &(0.041) &(0.068)& (0.068)&(0.306)  \\
\multicolumn{1}{l|}{\multirow{2}{*}{MCP}}                        &0.007          & 0.918 &  7.526    &  0.007 & 0.918     & 7.532\\
\multicolumn{1}{l|}{}                                & (0.003)&(0.060)  &(0.622) &(0.003)& (0.060)&(0.655)  \\
\multicolumn{1}{l|}{\multirow{2}{*}{STGP(no-info)}}                        &0.001          & 0.435 &  2.729    &  0.002 & 0.461     & 2.584\\
\multicolumn{1}{l|}{}                                & (0.002)&(0.161)  &(0.335) &(0.006)& (0.185)&(0.621)\\
\multicolumn{1}{l|}{\multirow{2}{*}{RCT}}                   &   0.025      &    0.018  &   2.302    &  0.027   &  0.196     & 3.038   \\
\multicolumn{1}{l|}{}                                 & (0.001)&(0.041) &(0.342) &(0.014)& (0.306)&(1.083)  \\ \hline
          & \multicolumn{3}{c|}{Model (7b)} & \multicolumn{3}{c}{Model (8b)}  \\ \hline
\multicolumn{1}{l|}{\multirow{2}{*}{Lasso}}          &0.002           & 0.825 &   7.328   &   0.001        &  0.805           &6.639   \\
\multicolumn{1}{l|}{}                                   & (0.002)& (0.051)&(1.719) &(0.001)& (0.053)&(0.429)    \\
\multicolumn{1}{l|}{\multirow{2}{*}{AdaLasso}}       & 0.003          &0.815   &  12.995   &  0.001   & 0.807      & 10.936   \\
\multicolumn{1}{l|}{}                               & (0.003)&(0.095) &(4.783) &(0.001)& (0.062)&(6.530)    \\
\multicolumn{1}{l|}{\multirow{2}{*}{MCP}}                        &0.007          & 0.918 &  7.525    &  0.007   & 0.918     & 7.532\\
\multicolumn{1}{l|}{}                                & (0.003)&(0.060)  &(0.619) &(0.007)& (0.060)&(0.654)  \\
\multicolumn{1}{l|}{\multirow{2}{*}{STGP(no-info)}}                        &0.001          & 0.460 &  2.904    &  0.001 & 0.485     & 2.600\\
\multicolumn{1}{l|}{}                                & (0.003)&(0.171)  &(0.635) &(0.006)& (0.192)&(0.650)  \\
\multicolumn{1}{l|}{\multirow{2}{*}{RCT}}                   & 0.030        &  0.053  &   2.520    &  0.030    & 0.172         & 2.949    \\
\multicolumn{1}{l|}{}                                & (0.007)& (0.033)&(0.219) &(0.013)& (0.265)&(0.828)  \\ \hline
          & \multicolumn{3}{c|}{Model (7c)} & \multicolumn{3}{c}{Model (8c)} \\ \hline
\multicolumn{1}{l|}{\multirow{2}{*}{Lasso}}           &   0.002             &0.854  &   7.228    &     0.041     &  0.418     &4.265 \\
\multicolumn{1}{l|}{}                                   & (0.001)& (0.033)&(1.222) &(0.006)& (0.050)&(0.360)   \\
\multicolumn{1}{l|}{\multirow{2}{*}{AdaLasso}}             & 0.001       & 0.853  &  10.690    &  0.033  & 0.446          & 4.347   \\
\multicolumn{1}{l|}{}                                 & (0.001)& (0.033)&(2.728) &(0.003)& (0.127)&(0.446)  \\
\multicolumn{1}{l|}{\multirow{2}{*}{MCP}}                        &0.007          & 0.918 &  7.524    &  0.007   & 0.918      & 7.535\\
\multicolumn{1}{l|}{}                                & (0.007)&(0.060)  &(0.612) &(0.003)& (0.069)&(0.675)  \\
\multicolumn{1}{l|}{\multirow{2}{*}{STGP(no-info)}}                        &0.001          & 0.484 &  2.753    &  0.003 & 0.476     & 2.561\\
\multicolumn{1}{l|}{}                                & (0.003)&(0.210)  &(0.610) &(0.008)& (0.226)&(0.577)  \\
\multicolumn{1}{l|}{\multirow{2}{*}{RCT}}          & 0.034         & 0.016  &   2.761   &  0.045    & 0.303         & 3.284  \\\multicolumn{1}{l|}{}                              & (0.008)& (0.105)   &(0.298) &(0.016)& (0.320)&(1.215)\\
\hline
\end{tabular}
}
\end{table}

As shown in Table 3,  Lasso, Adaptive Lasso, SCAD and MCP  fail to identify most of the true predictors and have a very high FNR, while our proposed estimator is consistently the best among all these models. It also indicates that the thresholding function helps us deal with these very complicated covariance structures of predictors. RCT also outperforms STGP(no-info), and has compatible performance with original STGP estimator, despite our estimator are not restricted with adjacency information.

We consider two more cases when the group lasso penalty is necessary to be applied. In Models 9--10, we partition the whole image space into 25 sub-regions with equal numbers of predictors in each region. The covariance structure inside each region is the same as Models 7--8. Correlations between different regions are 0.9. Then we randomly assign two regions with each non-zero regions of radius 0.13, which makes around 1/3 of the points within the selected region has non-zero effects. We assign all non-zero effects as 2. The covariance structures can be summarized as:
\begin{align*}
&\text{Model 9}\ (\text{GP25(10)}):\  \bbx \sim N(\bmu_b, \bSigma),\  \sigma_{ij}=\exp(-\|s_i\|^2-\|s_j\|^2-10\|s_i-s_j\|^2), \\
&\text{Model 10} \  (\text{GP25(5)}): \bbx \sim N(\bmu_b, \bSigma),\  \sigma_{ij}=\exp(-\|s_i\|^2-\|s_j\|^2-5\|s_i-s_j\|^2),
\end{align*}
where $\bmu_b$ is the mean for region $b$, and $(\bmu_1,\dots,\bmu_{25}) \sim N(0,\bGamma)$, $\bGamma_{ij}=0.9-0.1I(i=j)$. For the noise term, we still set $\sigma^2_1 = 2,4$ and $8$ as case (a),(b) and (c), respectively, and $\sigma^2_2=30$. For these two models, we compare performances of the Lasso, the group Lasso (GLasso), the sparse Group Lasso (SGL), STGP and the proposed estimator (RCT) in Table 4. As a key difference from Models $7$ and $8$,  Models $9$ and $10$ impose the group structure on both $\bbx$ and $\bbeta$. Figure 3(b) shows an illustration of the simulated imaging predictor $\bbx$ in Models 7 and 9.

\begin{table}
\center
\caption{Estimation and selection accuracy of different methods for Models 9 and 10}
{\footnotesize
\begin{tabular}{cccccc|ccccc}
\hline
                                                        &  FPR   &FNR  &R-FPR  &  R-FNR       &  $\ell_2$ loss   &  FPR   &FNR  &R-FPR  & R-FNR       &  $\ell_2$ loss  \\ \hline
                                                                                   & \multicolumn{5}{c|}{Model (9a)} & \multicolumn{5}{c}{Model (10a)}   \\ \hline
\multicolumn{1}{l|}{\multirow{2}{*}{Lasso}} & 0.019 & 0.270 & 0.101 & 0 & 25.607 & 0.028 & 0.411 & 0.140 & 0 & 33.091       \\
\multicolumn{1}{l|}{}                    &(0.004) & (0.072) & (0.091) & (0) & (3.235) & (0.006)& (0.100) & (0.083) & (0) & (3.380)   \\
\multicolumn{1}{l|}{\multirow{2}{*}{GLasso}} & 0.220 & 0.378 & 0.232 & 0 & 16.101 & 0.215 & 0.403 & 0.232 & 0 & 16.128\\
\multicolumn{1}{l|}{}             &(0.055) & (0.151) & (0.094) & (0) & (0.626) & (0.054) & (0.141) & (0.094) & (0) & (0.621) \\
\multicolumn{1}{l|}{\multirow{2}{*}{SGL}} &  0.115 & 0.010 & 0.135 & 0 & 12.507 & 0.126 & 0.012 & 0.126 & 0 & 13.366        \\
\multicolumn{1}{l|}{}               &(0.035) & (0.022) & (0.060) & (0) & (0.198) & (0.037) & (0.030) & (0.063) & (0) & (0.523)   \\
\multicolumn{1}{l|}{\multirow{2}{*}{STGP}} &  0.063 & 0 & 0.109 & 0 & 12.540 & 0.061 & 0 & 0.110 & 0 & 12.642      \\
\multicolumn{1}{l|}{}               &(0.010) & (0) & (0.060) & (0) & (0.281) & (0.007) & (0) & (0.060) & (0) & (0.354)   \\
\multicolumn{1}{l|}{\multirow{2}{*}{RCT}}  &0.059 & 0 & 0.087 & 0 & 13.114 & 0.064 & 0 & 0.111 & 0 & 13.505\\
\multicolumn{1}{l|}{}                     &(0.003) & (0) & (0.087) & (0) &(0.198) & (0.007) & (0) & (0.091) & (0) & (0.247)  \\ \hline
          & \multicolumn{5}{c|}{Model (9b)} & \multicolumn{5}{c}{Model (10b)}  \\ \hline
\multicolumn{1}{l|}{\multirow{2}{*}{Lasso}} &   0.019   & 0.347 & 0.166 & 0 & 27.557 & 0.028 & 0.415 & 0.145 & 0 & 32.957       \\
\multicolumn{1}{l|}{}        &(0.004) &(0.078) & (0.088) & (0) & (3.128) & (0.006) &(0) & (0.096) & (0) &(0.273)   \\
\multicolumn{1}{l|}{\multirow{2}{*}{GLasso}}   & 0.223 & 0.372 & 0.232 & 0 & 16.084 & 0.214 & 0.405 & 0.232 & 0 & 16.119\\
\multicolumn{1}{l|}{}                                & (0.053) & (0.143) & (0.094) & (0) & (0.618) & (0.056) & (0.140) & (0.094) & (0) & (0.619) \\
\multicolumn{1}{l|}{\multirow{2}{*}{SGL}} &   0.130 & 0.012 & 0.170 & 0 & 12.656 & 0.128 & 0.020 & 0.165 & 0 & 13.353       \\
\multicolumn{1}{l|}{}                     & (0.032) &(0.026) & (0.060) & (0) & (0.603) & (0.037) & (0.046) & (0.091) & (0) & (0.615)   \\
\multicolumn{1}{l|}{\multirow{2}{*}{STGP}} &  0.061 & 0 & 0.114 & 0 & 12.520 & 0.062 & 0 & 0.113 & 0 & 12.536        \\
\multicolumn{1}{l|}{}               &(0.007) & (0) & (0.062) & (0) & (0.378) & (0.007) & (0) & (0.039) & (0) & (0.173)   \\
\multicolumn{1}{l|}{\multirow{2}{*}{RCT}}  &0.066 & 0 & 0.161 & 0 & 13.269 & 0.065 & 0 & 0.120 & 0 & 13.670\\
\multicolumn{1}{l|}{}         &(0.006) & (0) & (0.100) & (0) & (0.281) & (0.006) & (0) & (0.096) & (0) & (0.273) \\ \hline
          & \multicolumn{5}{c|}{Model (9c)} & \multicolumn{5}{c}{Model (10c)} \\ \hline
\multicolumn{1}{l|}{\multirow{2}{*}{Lasso}} &   0.019     &  0.394 & 0.180 & 0 & 27.949 & 0.027 & 0.427 & 0.151 & 0 & 32.553      \\
\multicolumn{1}{l|}{}                   &(0.004) &(0.075) & (0.102) & (0) & (3.328) & (0.005) & (0.098) & (0.098) & (0) & (3.517)  \\
\multicolumn{1}{l|}{\multirow{2}{*}{GLasso}}  & 0.223 & 0.354 & 0.232 & 0 & 16.064 & 0.211 & 0.407 & 0.232 & 0 & 16.114\\
\multicolumn{1}{l|}{}         &(0.058) &(0.143) &(0.094) & (0) & (0.619) & (0.054) & (0.137) & (0.094) &(0) & (0.634) \\
\multicolumn{1}{l|}{\multirow{2}{*}{SGL}} & 0.128 & 0.013 & 0.165 & 0 & 12.869 &0.135 & 0.016 & 0.213 & 0 & 13.566      \\
\multicolumn{1}{l|}{}      &(0.033) &(0.020) &(0.123) & (0) & (0.700) &(0.040) & (0.027) & (0.170) & (0) & (0.848)  \\
\multicolumn{1}{l|}{\multirow{2}{*}{STGP}} &  0.059 & 0 & 0.098 & 0 & 12.565 & 0.126 & 0.012 & 0.126 & 0 & 13.366        \\
\multicolumn{1}{l|}{}               &(0.006) & (0) & (0.064) & (0) & (0.362) & (0.037) & (0.030) & (0.063) & (0) & (0.523)   \\
\multicolumn{1}{l|}{\multirow{2}{*}{RCT}} & 0.067 & 0 & 0.157 & 0 & 13.581 &0.059 & 0 &0.088 & 0 & 12.655 \\
\multicolumn{1}{l|}{}       &(0.006) & (0) & (0.092) & (0) & (0.260) & (0.007) & (0) & (0.066) & (0) & (0.324) \\ \hline
\end{tabular}
}

\end{table}

\begin{figure}
	\centering
	\subfigure[Model 7]{\includegraphics[width=2.4in]{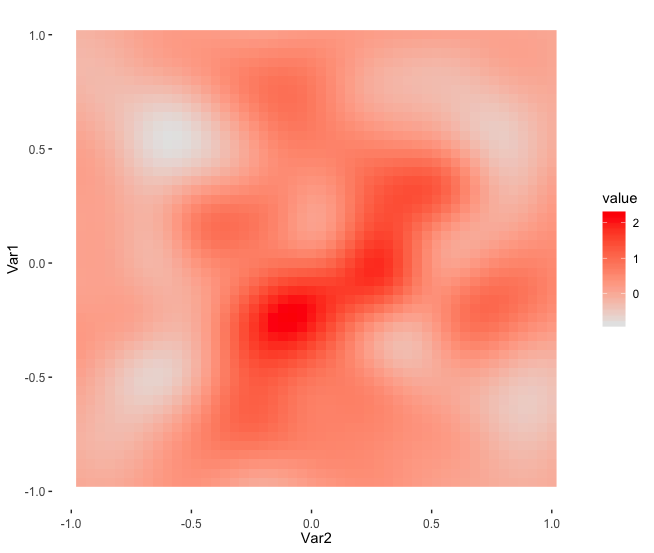} }
	\subfigure[Model 9]{\includegraphics[width=2.4in]{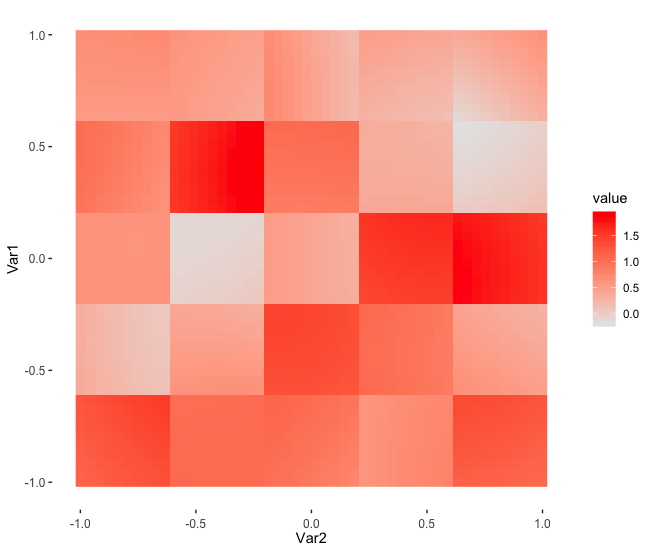} }
	\caption{Plots of the simulated imaging predictors from the Gaussian process regression}\label{fig:2}
\end{figure}
In Table 4, we include the region level false positive rate~(Region FPR) and region level false negative rate~(region FNR) to measure the variable selection accuracy within each region. They are computed based on whether there is at least one variable in the region is selected. Compared with the group $\ell_1$ penalty, the proposed method identified almost all the correct groups with zero false negatives and achieved a more satisfying FPR and the $\ell_2$ loss. The proposed method is less likely to over-select either predictors or regions than the sparse group Lasso. It implies that the thresholding function effectively achieved a lower FPR. Again, RCT has compatible performance with original STGP estimator, despite our estimator are not restricted with adjacency information.

\section{Application to Scalar-on-Image Analysis}\label{section 5}

This section applies the proposed method to analyze the 2-back versus 0-back contrast maps derived from the $n$-back task fMRI imaging data in the Adolescent Brain Cognitive Development~(ABCD) study~\citep{casey2018adolescent}.  Our goal is to identify the important imaging features from the contrast maps that are strongly associated with the risk of psychairtric disorder, measured by the general factor of psychopathology~(GFP) or ``p-factor".  After the standard fMRI prepossessing steps, all the images are registered into the 2 mm standard Montreal Neurological Institute (MNI) space consisting of 160,990 voxels in the 90 Automated Anatomical Labeling (AAL) brain regions.  
With the missing values being removed, the data used in our analysis consists of 2,070 subjects. To reduce the dimension of the imaging data, we partition 90 AAL regions into 2,518  sub-regions with each region consisting of an average of 64 voxels. We refer to each subregion as a super-voxel. For each subject, we compute the average intensity values of the voxels within each super-voxel as its intensity. We consider those 2,518 super-voxel-wise intensity values as the potential imaging predictors. 

There are several challenging issues in the scalar-on-image regression analysis of this dataset. First, the correlations between  super-voxels across the 90 AAL regions can be very high and the correlation patterns are complex. In fact, there are 151,724 voxel pairs across these regions having a correlation larger than 0.8~(or less than $-0.8$), and 9,038 voxel pairs with a correlation larger than 0.9~(or less than $-0.9$). Figure \ref{cor_illustration} visualizes the region-wise correlation structures, where panel (a) shows the highest correlations between regions; and panel (b) counts the voxel pairs that have a correlation higher than 0.8~(or less than $-0.8$) in each corresponding region pair.  Given the imaging predictors having such high and complicated covariance structures, the classical lasso or the group lasso method may fail to perform variable selection satisfactorily. In contrast, the proposed model with coefficient thresholding is developed to resolve this issue since it does not require the strong conditions on the design matrix. Second, the AAL brain atlas provides useful information on the brain structure and function that may be related to the risk of psychiatry disorder. It is of interest to integrate the AAL region partition as grouping information of imaging predictors to improve the accuracy of imaging feature selection. Third, the outcome variable ``p-factor" has a right skewed marginal distribution (see Figure 4), thus its measurements are more likely to contain outliers.  The existing non-robust scalar-on-image regression methods may produce inaccurate results. 
 All the aforementioned challenging issues motivate the needs of developing our  robust regression with coefficient thresholding and group penalty.

\begin{figure}\label{p_factor}
\center
 \includegraphics[width=2.3in]{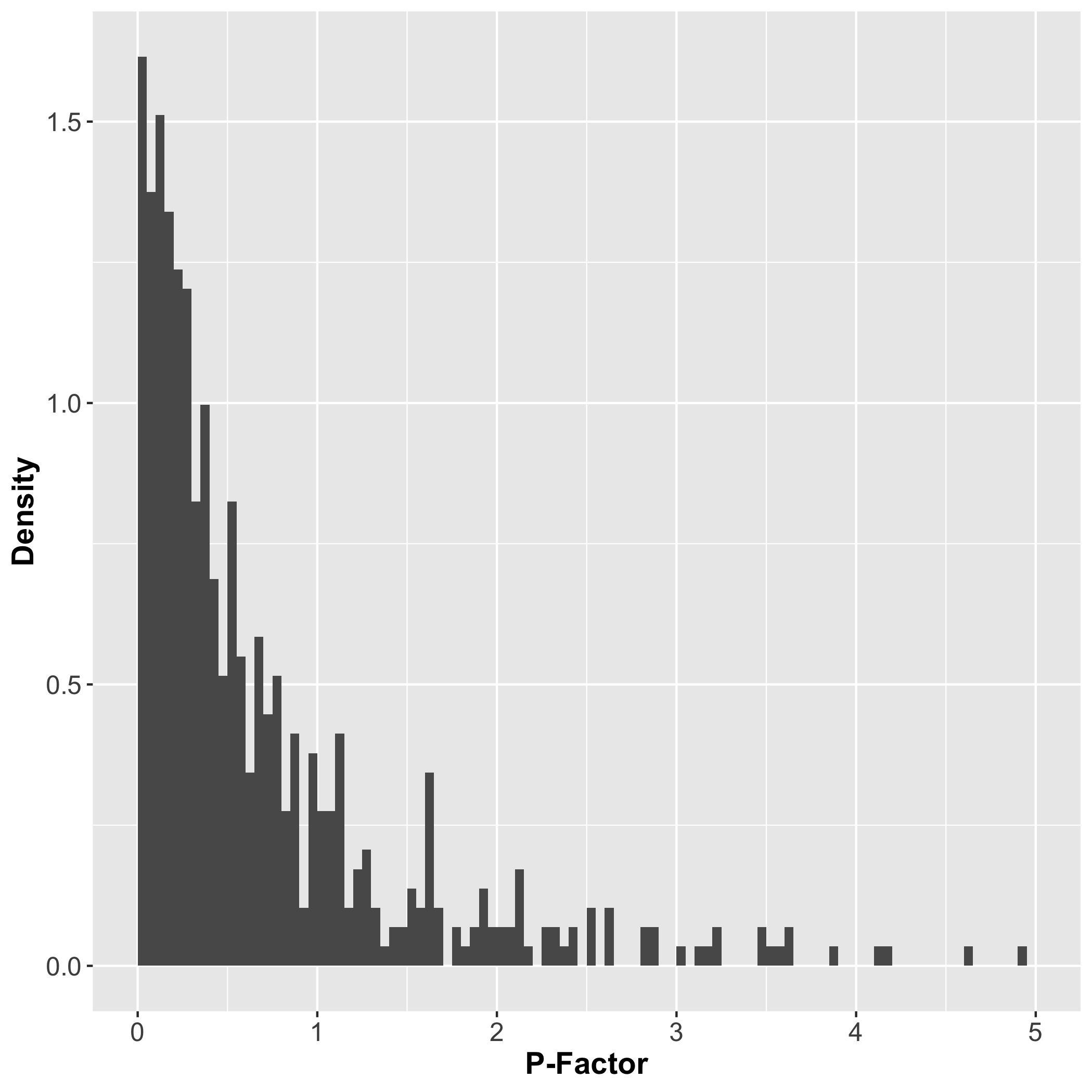}
	\caption{Marginal distribution of `p-factor' variable among all subjects}
\end{figure}

In our analysis, we adjust confounding effects by including a few predictors in the model: family size, gender, race, highest parents education, household marital status and household income level. Given the intrinsic group structure, we compare the performance of the proposed method and the STGP in this data analysis. 
As we mentioned, the fMRI data analysis generally suffers from the reliability issue due to its complex data structure and low signal-to-noise ratio~\citep{bennett2010reliable,brown2017controversy,eklund2012does,eklund2016cluster}. To evaluate the variable selection stability for both methods, we consider a bootstrap approach with 100 replications. In each replication, we sample $n$ observations with replacement, and fit the bootstrap samples using the best set of  tuning parameters chosen by a five-fold cross-validation. Then we obtain the frequency of each super-voxel being selected over 100 replications as a measure of the selection stability, which can be used to fairly compare the regions that can be consistently selected against randomness, and thus ensure the reliability of the scientific findings in our analysis.

\begin{figure}[H]
	\centering
		\subfigure[Soft Thresholded Gaussian Process]{\includegraphics[width=2.3in]{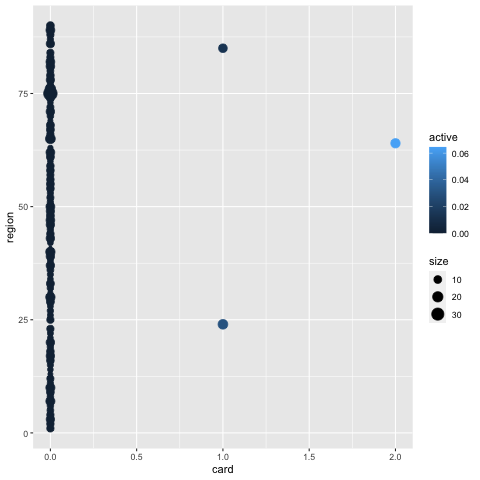} }
	\subfigure[Robust Coefficient Thresholding]{\includegraphics[width=2.3in]{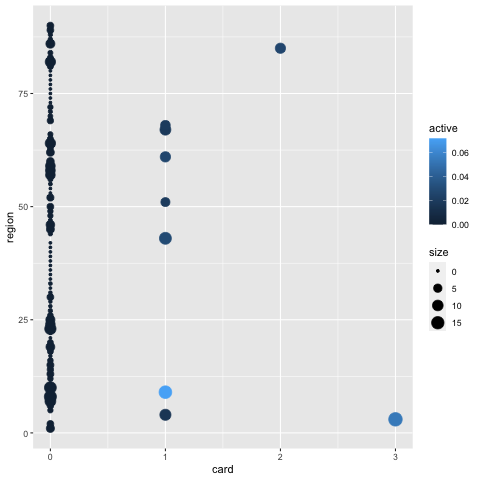} }
	\caption{Illustration of the selection frequency and cardinality of 90 brain regions for STGP and the RCT. The X axis shows the maximum selection frequency, the bubble size is proportional to the number of super-voxels with the selection frequency being larger than $0.6$, and the color indicates the proportion of super-voxels for each brain region.}\label{fig:4}
\end{figure}

RCT and STGP respectively select  124.5 and 245.3 super-voxels per replication on average. Figure \ref{fig:4} displays the bootstrap selection results, where the $x$-axis represents the maximum selection frequency of super-voxels in each region. The circle size is proportional to the number of  super-voxels with the corresponding selection frequency being larger than 0.6. The color represents the proportion of super-voxels being selected in each region. 
Despite a smaller number of super-voxels being selected in each bootstrap run, RCT consistently selects super-voxels in several important brain regions over bootstrap samples, while STGP identifies a less number of brain regions that contain selected super-voxels.

\begin{table}[H]\label{table5}
\center
\caption{Comparisons of stable selection regions between RCT and STGP for different levels of selection frequency threshold~\label{tab:IQ}}
\smallskip
\footnotesize{
\begin{tabular}{l|l|l}
\hline
Selected frequency       & RCT         & STGP                                                                          \\ \hline
\multirow{2}{*}{0.6-0.7} & Callcarine\_L, Occipital\_Mid\_L,  & Frontal\_Sup\_Midial\_R, Temporal\_Mid\_L           \\
                         & Parietal\_Inf\_L                   &                               \\ \hline
0.7-0.8                  & Temporal\_Mid\_L     & SupraMarginal\_R                            \\          \hline
0.8-0.9                  & Frontal\_Mid\_Orb\_L, Precuneus\_L & N/A                                    \\ \hline
\textgreater{}0.9        & Front\_Sup\_L                      & N/A                                                                        \\
\hline
\end{tabular}
}
\end{table}

Table \ref{table5} summarizes the comparisons of selected regions from RCT and STGP by varying different thresholds of selection frequency from 0.6 to 0.9. Compared with STGP, RCT selects more stable regions for each level of selection frequency, indicating that our method produces more reliable selection results.  In particular, containing at least one super-voxels with more than 60\% selection frequency, seven and three regions are respectively identified by RCT and STGP. Among those regions, only one common region, i.e. the left temporal gyrus (Temporal\_Mid\_L), is detected by both methods, where RCT has a higher selection frequency (0.79) than STGP (0.69). The existing functional neuroimaging studies have indicated that the middle temporal gyrus is involved in language and semantic memory processing~\citep{cabeza2000imaging}, and it is also related to mental diseases such as chronic schizophrenia~\citep{onitsuka2004middle}. 

Among other selected regions, with more than 90\% selection frequency, RCT consistently selects super-voxels in the left superior frontal gyrus (Frontal\_Sup\_L), while the selection frequency by STGP is below 60\%. Superior frontal gyrus is known to be strongly related with working memory~\citep{boisgueheneuc2006functions} which plays a critical role in attending to and analyzing incoming information. Deficits in working memory are associated with many cognitive and mental health challenges, such as anxiety and stress~\citep{lukasik2019relationship}, which can be captured by the ``p-factor". The strong relationship between p-factor and working memory has been discovered by exisitng studies~\citep{huang2017poor}. 

In addition, RCT also identifies five more regions than STGP:  precuneus, the left middle frontal gyrus (Frontal\_Mid\_Orb\_L), the left alcarine fissure and surrounding cortex (Calcarine\_L), the middle occipital gyrus (Occipital\_Mid\_L) and the left inferior parietal gyri (Parietal\_Inf\_L). Percuneus is well studied as a core of mind~\citep{cavanna2006precuneus}, and it is highly related to posttraumatic stress disorder(PTSD) and other mental health issue~\citep{geuze2007precuneal}. Middle frontal gyrus is part of limbic system and known to be highly related to emotion~\citep{sprooten2017addressing}. Inferior parietal lobule has been involved in the perception of emotions in facial stimuli, and interpretation of sensory information~\citep{radua2010neural}. Calcarine fissure is related to vision.  Middle occipital gyrus is primarily responsible for object recognition. It would be interesting to further investigate how the brain activity in these regions influence on the p-factor.

To further demonstrate the proposed method providing more reliable scientific findings in comparison to  STGP, we evaluate the prediction performance of the two methods by using the train-test data cross-validation. We randomly split the data into two parts with $80\%$ as the training data for model fitting  and  $20\%$ as the test data for computing the prediction error. We repeat this procedure for 50 times. The mean absolute prediction error of RCT is 0.464 with standard error 0.004; STGP has a mean absolute prediction error of 0.480 with standard error 0.038. We conclude that our proposed method improves the prediction performance of the p-factor using working memory contrast maps in the ABCD study, comparing with the state-of-the-art scalar-on-image regression method.

\section{Conclusion}\label{section 6}

In this paper, we propose a novel high-dimensional robust regression with coefficient thresholding in the presence of complex dependencies among predictors and potential outliers. The proposed method uses the power of thresholding functions and the robust Huber loss to build an efficient nonconvex estimation procedure. We carefully analyze the landscape of the nonconvex loss function for the proposed method, which enables us to establish both statistical and computational consistency. We demonstrate the effectiveness and usefulness of the proposed method in simulation studies and a real application to imaging data analysis. { In the future, it is interesting to investigate how to incorporate the spatial-temporal information of the imaging data into our proposed method. It is also important to study the statistical consistency of the near-stationary solution from the proposed gradient descent based algorithm under more general conditions. }

\bibliographystyle{agsm}
\bibliography{threg}

\end{document}